\documentclass[10pt,a4paper,onecolumn]{article}                

\usepackage{latexsym}
\usepackage[a4paper,left=2.5cm,right=2.5cm,top=2.5cm,bottom=2.5cm]{geometry}
\usepackage{amsmath}
\usepackage{amsfonts}
\usepackage[pagebackref]{hyperref}
\begin{document}

\title{The statistical origins of quantum mechanics}

\author{U. Klein\thanks{ulf.klein@jku.at}\\Johannes Kepler Universit{\"a}t Linz\\Institut f{\"u}r Theoretische Physik\\ A-4040 Linz, Austria\\}

\date{\today}

\maketitle

\bibliographystyle{plain}

\begin{abstract}
It is shown that Schr{\"o}dinger's equation may be derived from 
three postulates. The first is a kind of statistical metamorphosis 
of classical mechanics, a set of two relations which are obtained 
from the canonical equations of particle mechanics by replacing all 
observables by statistical averages. The second is a local 
conservation law of probability with a probability current which 
takes the form of a gradient. The third is a principle of 
maximal disorder as realized by the requirement of minimal Fisher
information. The rule for calculating expectation values is 
obtained from a fourth postulate, the requirement of energy 
conservation in the mean. The fact that all these basic relations 
of quantum theory may be derived from premises which are statistical 
in character is interpreted as a strong argument in favor of the 
statistical interpretation of quantum mechanics. The structures of 
quantum theory and classical statistical theories are compared and 
some fundamental differences are identified.
\end{abstract}


\section{Introduction}
\label{sec:introduction}
The interpretation of quantum theory does neither influence its
theoretical predictions nor the experimentally observed data. 
Nevertheless it is extremely important because it determines 
the direction of future research. One of the many controversial 
interpretations  of quantum mechanics is the ``statistical interpretation'' 
or ``ensemble interpretation''~\cite{ballentine:statistical}. It 
presents a point of view, which is in opposition to most variants 
of the Copenhagen interpretation~\cite{barrett:quantum}, 
but has been advocated by a large number of eminent physicists, 
including Einstein. It claims that quantum mechanics is incomplete with 
regard to the description of single events and that all its dynamic 
predictions are of a purely statistical nature. This means that, in general, 
a large number of measurements on identically prepared systems have to be 
performed in order to verify a (dynamical) prediction of quantum theory. 

The origin of the time-dependent Schr{\"o}dinger equation is of course
an essential aspect for the interpretation of quantum mechanics.
Recently a number of derivations of Schr{\"o}dinger's equation have been 
reported which use as a starting point not a particle Hamiltonian but 
a statistical ensemble. The basic assumptions underlying these works include
special postulates about the structure of momentum 
fluctuations~\cite{hall.reginatto:schroedinger}, the principle of 
minimum Fisher information~\cite{reginatto:derivation,syska:fisher}, a 
linear time-evolution law for a complex state 
variable~\cite{klein:schroedingers}, or the assumption of a 
classical stochastic force of unspecified 
form~\cite{kaniadakis:statistical}. The work reported in this paper
belongs to this class of theories, which do not ``quantize'' a single 
particle but a statistical ensemble. An attempt is undertaken to 
improve this approach by starting from assumptions, which may be
considered as simpler and more fundamental from a physical point of view.
It is shown that Schr{\"o}dinger's equation may be derived from a small 
number of very general and simple assumptions - which are all essentially of 
a statistical nature. In a first step an infinite class of statistical theories is 
derived, containing a classical statistical theory as well as quantum mechanics. 
In a second step quantum mechanics is singled out as ``most reasonable 
statistical theory'' by imposing as an additional requirement the principle of 
maximal disorder, as realized by the principle of minimal Fisher information. 

We begin in section~\ref{sec:onprobability} with a general discussion 
of the role of probability in physical theories. In 
section~\ref{sec:statisticalconditions} the central 'statistical 
condition' (first assumption) of this work is formulated.
The set of corresponding statistical theories is derived in 
section~\ref{sec:statisticaltheories}. In 
sections~\ref{sec:randomvariables} and~\ref{sec:disorder} 
structural differences between quantum theory and classical 
statistical theories are investigated. The quantum mechanical rule 
for calculating expectation values is derived from the 
requirement of conservation of energy in the mean in 
section~\ref{sec:energyconservation}. 
In sections~\ref{sec:disorder}-\ref{sec:subsidiary-condition}
the principle of maximal disorder is implemented and Fisher's 
information measure is derived in section~\ref{sec:derivationofL}. 
Section~\ref{sec:discussion} contains a detailed discussion of all 
assumptions and results and may be consulted in a first reading 
to obtain an overview of this work; questions of interpretation of 
the quantum theoretical formalism are also discussed in this section. 
In the last section~\ref{sec:concludingremarks} open questions for 
future research are listed.

\section{On probability}
\label{sec:onprobability}
With regard to the role of probability, three types of physical 
theories may be distinguished.
\begin{itemize}
\item [1.] Theories of type 1 are deterministic. Single events are 
completely described by their known initial values  and deterministic 
laws (differential equations). Classical mechanics is obviously such a 
theory. We include this type of theory, where probability does not 
play any role, in our classification scheme  because it provides a 
basis for the following two types of theories.
\item [2.] Theories of type 2 have deterministic laws but the initial 
values are unknown. Therefore, no predictions on individual events are 
possible, despite the fact that deterministic laws describing 
individual events are valid. In order to verify a prediction of a 
type 2 theory a large number of identically prepared experiments must 
be performed. We have no problems to understand or to interpret such a 
theory because we know its just our lack of knowledge which causes the 
uncertainty. An example is given by classical statistical mechanics. 
Of course, in order to construct a type 2 theory one needs a type 1 
theory providing the deterministic laws.  
\item [3.] It is possible to go one step further in this direction  
increasing the relative importance of probability even more. We may 
not only work with unknown initial values but with unknown laws as 
well. In these type 3 theories there are no deterministic laws describing 
individual events, only probabilities can be assigned. There is no 
need to mention initial values for particle trajectories any more
(initial values for probabilistic dynamical variables are still 
required).
\end{itemize}
Type 2 theories could also be referred to as classical (statistical) 
theories. Type 3 theories are most interesting because we recognize 
here characteristic features of quantum mechanics. In what follows we 
shall try to make this last statement more definite.

Comparing type 2 and type 3 theories, one finds two remarkable aspects. 
The first is a subtle kind of ``inconsistency'' of type 2 theories: If 
we are unable to know the initial values of our observables (at a particular 
time), why should we be able to know these values during the following 
time interval (given we know them at a fixed time). In other words, in 
type 2 theories the two factors determining the final outcome of a 
theoretical prediction - namely initial values and laws - are not placed 
on the same (realistic) footing. This hybrid situation has been 
recognized before; the term 'crypto-deterministic' has been 
used by Moyal~\cite{moyal:quantum} to characterize classical statistical 
mechanics (note that the same term is also used in a very different sense 
to characterize hidden variable theories~\cite{peres:quantum}). Type 3 
theories do not show this kind of inconsistency.

The second observation is simply that type 2 and type 3 theories have 
a number of important properties in common. Both are unable to 
predict the outcome of single events with certainty; only probabilities
are provided in both cases. In both theories the quantities which may 
be actually observed  - whose time dependence may be formulated in terms 
of a differential equation - are averaged observables, obtained with the 
help of a large number of single experiments. These common features lead 
us to suspect that a general structure might exist which comprises 
both types of theories.  

Such a general structure should consist of a set of (statistical) 
conditions, 
which have to be obeyed by any statistical theory. In theories of this 
kind  observables in the conventional sense do not exist. Their 
role is  taken over by  random variables. Likewise, conventional physical 
laws - differential equations for time-dependent observables - do not exist. 
They are replaced by differential equations for statistical averages. These 
averages of the (former) observables become the new observables, with the 
time $t$ playing again the role of the independent variable. In order 
to construct such general conditions one needs again (as with type 2 
theories) a deterministic (type I) theory as a ``parent'' theory. 
Given such a type 1 theory, we realize that a simple recipe to construct 
a reasonable set of statistical conditions is the following: Replace 
all observables (of the type 1 theory) by averaged values using 
appropriate probability densities. In this way the dynamics of the 
problem is completely transferred from the observables to the 
probability distributions. This program will be carried through 
in the next sections, using a model system of classical mechanics as 
parent theory.  

The above construction principle describes an unusual situation, because 
we are used to considering determinism (concerning single events) as a 
very condition for doing science. Nevertheless, the physical context, 
which is referred to is quite simple and clear, namely that nature forbids  
for some reason deterministic description of single events but allows 
it at least ``on the average''. It is certainly true that we are not 
accustomed to such a kind of thinking. But to believe or not to believe 
in such mechanisms of nature is basically a matter of intellectual habit. 
Also, the fact that quantum mechanics is incomplete does not necessarily 
imply that a complete theory exists; the opposite possibility, that no 
deterministic description of nature will ever be found, should also be 
taken into account. 

\section{Statistical conditions}
\label{sec:statisticalconditions}
We study a simple system, a particle in an externally controlled 
time-independent potential $V(x)$, whose motion is restricted to a 
single spatial dimension (coordinate $x$). We use the canonical formalism 
of classical mechanics to describe this system. Thus, the fundamental 
observables of our theory are  $x(t)$ and  $p(t)$ and they obey the 
differential equations  
\begin{equation}
\label{eq:FSTASECCANEQ}
\frac{\mathrm{d}}{\mathrm{d}t} x(t)  =  \frac{p(t)}{m},
\hspace{0.5cm}
\frac{\mathrm{d}}{\mathrm{d}t} p(t)  =   F(x(t)) 
\mbox{,}
\end{equation}
where $F(x)=-\frac{\mathrm{d}V(x)}{\mathrm{d}x}$. We now create 
statistical conditions, associated with the type 1 
theory~(\ref{eq:FSTASECCANEQ}), according to the method outlined in 
the last section. We replace the observables $x(t),\,p(t)$ and the 
force field $F(x(t))$ by averages $\overline{x},\,\overline{p}$ 
and $\overline{F}$, and obtain
\begin{eqnarray}
\frac{\mathrm{d}}{\mathrm{d}t} \overline{x} & = & \frac{\overline{p}}{m}
\label{eq:FIRSTAET}\\
\frac{\mathrm{d}}{\mathrm{d}t} \overline{p} & = & \overline{F(x)} 
\label{eq:SECONAET}
\mbox{,}
\end{eqnarray}
The averages in~(\ref{eq:FIRSTAET}),(\ref{eq:SECONAET}) are mean 
values of the random variables $x$ or $p$; there is no danger of 
confusion here, because the symbols $x(t)$ and $p(t)$ will not be 
used any more. In~(\ref{eq:FSTASECCANEQ}) only terms occur, which 
depend either on the coordinate or the momentum, but not on both. Thus, 
to form the averages we need two probability densities $\rho(x,t)$ and 
$w(p,t)$, depending on the spatial coordinate $x$ and the momentum $p$ 
separately. Then, the averages occurring 
in~(\ref{eq:FIRSTAET}),(\ref{eq:SECONAET}) are given by  
\begin{eqnarray}
\overline{x} & = & \int_{-\infty}^{\infty} \mathrm{d} x \rho(x,t) x 
\label{eq:ERWAETX}\\
\overline{p} & = & \int_{-\infty}^{\infty} \mathrm{d} p w(p,t) p 
\label{eq:ERWAETP}\\
\overline{F(x)} & = & - \int_{-\infty}^{\infty} \mathrm{d} x \rho(x,t)
\frac{\mathrm{d}V(x)}{\mathrm{d}x}
\label{eq:ERWAETFX}
\mbox{.}
\end{eqnarray}
Note that $F(x)$ has to be replaced by $\overline{F(x)}$ and not by 
$F(\overline{x})$. The probability densities $\rho$ and $w$ are positive 
semidefinite and normalized to unity. They are time-dependent because 
they describe the dynamic behavior of this theory. 

Relations~(\ref{eq:FIRSTAET}),(\ref{eq:SECONAET}), with the 
definitions~(\ref{eq:ERWAETX})-(\ref{eq:ERWAETFX}) are, to the best
of my knowledge, new. They will be referred to as ``statistical conditions''.
There is obviously a formal similarity 
of~(\ref{eq:FIRSTAET}),(\ref{eq:SECONAET}) 
with Ehrenfest's relations of quantum mechanics, but the differential 
equations to be fulfilled by $\rho$ and $w$ are still unknown and may well 
differ from those of quantum theory. 
Relations~(\ref{eq:FIRSTAET})-(\ref{eq:ERWAETFX}) represent general 
conditions for theories which are deterministic only with respect to 
statistical averages of observables and not with respect to single events. 
They cannot be associated to either the classical or the quantum mechanical 
domain of physics.  Many concrete statistical theories (differential 
equations for the probability distributions) obeying these conditions may 
exist (see the next section). 

These conditions should be supplemented by a local conservation law of 
probability. Assuming that the probability current is proportional to 
the gradient of a function $S$ (this is the simplest possible choice 
and the one realized in Hamilton-Jacobi theory, see also 
section~\ref{sec:discussion}) this conservation law is 
for our one-dimensional situation given by the continuity equation
\begin{equation}
  \label{eq:CONTONEDIM}
\frac{\partial \rho(x,t)}{\partial t}+   
\frac{\partial}{\partial x} \frac{\rho(x,t)}{m}\frac{\partial S(x,t)}{\partial x}=0
\mbox{.}
\end{equation} 
The derivative of $S(x,t)$ defines a field with dimension of a momentum, 
\begin{equation}
  \label{eq:DEFMOMD1}
p(x,\,t)=\frac{\partial S(x,\,t)}{\partial x}
\mbox{.}
\end{equation}
Eq.~\eqref{eq:DEFMOMD1} defines a unique number $p(x,\,t)$ for each 
value of the random variable $x$. In the next section we will discuss 
the following question: Are we allowed to identify 
the possible values of the random variable  $p$ occurring in 
Eq.~\eqref{eq:ERWAETP} with the values of the momentum field $p(x,\,t)$ ?

\section{On random variables}
\label{sec:randomvariables}
Introducing standard notions of probability theory, the fundamental 
sample space of the present theory is given by all possible results of 
position measurements, i.e. it may be identified with the set of real numbers 
$\mathbb{R}$. This set $\mathbb{R}$ may also be identified with the possible 
values of a random variable ``position measurement'' (whose name should 
strictly speaking differ from $x$ but we shall neglect such differences here). 
The basic probability measure which assigns a probability to each event 
(subspace of $\mathbb{R}$) is given by $\rho(x,\,t)$. According to 
standard probability theory the field $p(x,\,t)$ defined 
by~\eqref{eq:DEFMOMD1} is itself a random variable. We may consider it as a
function of the random variable $X$ (denoting ``position measurement'') or 
as a random variable defined independently on the fundamental 
event space $\mathbb{R}$; it makes no difference. Its probability 
density is uniquely determined by $\rho(x,\,t)$ and the function $p(x,t)$.
In order to avoid confusion of names it may be useful to denote the 
derivative of $S(x,\,t)$ with respect to $x$ by $g(x,t)$ instead 
of $p(x,t)$. Thus $p(x,t)=g(x,t)$ and the notation $p=g(x,t)$ 
indicates that a random variable $p$ defined by the function $g(x,t)$ 
exists (the time variable will sometimes be omitted for brevity). 

In order to study this important point further, we rewrite the 
standard result for the probability density of $p=g(x,t)$ 
in a form more appropriate for physical considerations (a form 
apparently not easily found in textbooks on probability). 
For the simplest possible situation, a denumerable sample space 
with elements $x_i$, a probability measure $P$, and a 
invertible function $g(x)$, the probability that an event 
$p_i=g(x_i)$ occurs is obviously given by $W(p_i)=P(g^{-1}(p_i))$. 
This result is the starting point to obtain $w(p)$, the probability 
density  of a continuous random variable $p=g(x)$, which is defined by
a non-invertible function $g(x)$. It is given by~\cite{soong:probability} 
\begin{equation}
  \label{eq:aesf4fp}
w(p)=\sum_{i=1}^{n(p)}\rho(g_i^{-1}(p))
\left| \frac{\partial g_i^{-1}(p)}{\partial p} \right|
\mbox{,}
\end{equation}
where $g_i^{-1}(p)$ denotes the $n(p)$ solutions (the number of solutions 
depends on $p$) of the equation $p-g(x)=0$. Using a well-known 
formula for Dirac`s delta function $\delta$, applied to the case where 
the argument of $\delta$ is an arbitrary function, Eq.\eqref{eq:aesf4fp} 
may be rewritten in the form 
\begin{equation}
  \label{eq:asff8eq}
w(p,t)=\int\,\mathrm{d}x\,\rho(x,t)\delta\left(p-\frac{\partial S(x,t)}{\partial x} \right)
\mbox{,} 
\end{equation}
where we came back to our original notation, writing down the 
$t-$dependencies of $\rho$ and $w$ and replacing $g(x)$ by $\partial S(x,t)/\partial x$. 

The representation~\eqref{eq:asff8eq} reveals very clearly a hybrid 
nature of random variables defined as (nontrivial) functions on 
the event space $\mathbb{R}$. They are partly defined by a probabilistic 
quantity [namely $\rho(x)$] and partly by a deterministic relation 
[namely $g(x)$]. The deterministic nature of the latter is expressed 
by the singular (delta-function) shape of the associated probability. 
Such densities occur in classical statistics, i.e. in type 2 theories; 
Eq.~\eqref{eq:asff8eq} may obviously be obtained by performing an 
integration over $x$ of the classical phase space probability 
density $\rho(x,t)\delta(p-\partial S(x,t)/\partial x)$. 
Considered from an operational point of view, the hybrid nature of 
random variables may be described as follows. Deterministic predictions
for random variables $p=g(x)$ are impossible, as are deterministic 
predictions for the original variables $x$. But once a number $x$ has 
been observed in an experiment, then the value of $p=g(x)$ is 
\emph{with certainty} given by the defining function $g(x)$. If 
no such relation exists, this does not necessarily imply that    
$x$ and $p$ are completely independent. Many other more complicated 
('nonlocal' or 'probabilistic') relations between such variables are 
conceivable.    

We formulated general conditions comprising both type 2 and type 3 theories. 
Thus, as far as this general framework is concerned we can certainly not 
dispense with the standard notion of random variables, which are basic 
ingredients of type 2 theories; such variables will certainly occur as 
special (type 2) cases in our formalism. But, of course, we are essentially 
interested in the characterization of type 3 theories and the form of 
Eq.~\eqref{eq:asff8eq} shows that the standard notion of random variable 
is \emph{not} necessarily meaningful in a type 3 theory. Thus we will 
\emph{allow} for the possibility of random variables which are not 
defined by deterministic relations of the standard type, as functions 
on the sample space. 

This situation leads to a number of questions. We may, e.g. ask: Can 
we completely dispense with the the standard concept of random variables 
if we are dealing exclusively with a type 3 theory ? The answer is certainly 
no; it seems impossible to formulate a physical theory without any 
deterministic relations. In fact, a deterministic relation, corresponding 
to a standard random variable $F(x)$, has already been anticipated in 
Eq.~(\ref{eq:ERWAETFX}). If in a position measurement of a particle a 
number $x$ is observed, then the particle is - at the time of the 
measurement - \emph{with certainty} under the influence of a force $F(x)$. 
Thus, an allowed class of deterministic relations might contain ``given'' 
functions, describing externally controlled influences like forces $F(x)$ 
or potentials $V(x)$. 

There may be other standard random variables. To decide on purely logical 
grounds which relations of a type 3 theory are deterministic and which are 
not is not an obvious matter. However, one would suspect that the 
deterministic relations should be of an universal nature; e.g. they should 
hold both in type 2 and type 3 theories. Further, we may expect that all 
relations which are a logical consequence of the structure of space-time 
should belong to this class. Such a quantity is the kinetic energy. In 
fact, for the currently considered nonrelativistic range of physics, 
the functional form of the kinetic energy can be derived from the 
structure of the Galilei group both in the mathematical framework 
of classical mechanics~\cite{jordan:how_relativity} and quantum 
mechanics~\cite{jordan:why_momentum}. We refer to the kinetic 
energy $p^{2}/2m$ as a standard random variable insofar as it is a 
prescribed function of $p$ (but it is, because it is a function of $p$, 
not a standard random variable with respect to the fundamental 
probability measure $\rho$). Combining the standard random variables 
``kinetic energy'' and ``potential'' we obtain a standard random 
variable ``energy'', which will be studied in more detail in 
section~\ref{sec:energyconservation}. 

Thus, in the present framework, particle momentum will, in general, 
not be considered as a standard random variable. 
This means that an element of determinism has been eliminated from 
the theoretical description. It seems that this elimination is one 
of the basic steps in the transition from type 2 to type 3 theories. 
The functional form of the probability density $w(p,t)$, and its 
relation to $\rho(x,t)$, are one of the main objectives of the present 
study. According to the above discussion a measurement of position 
does no longer determine momentum at the time of the measurement. 
However the set of all position measurements [represented formally 
by the probability density $\rho (x,t)$] may still determine (in a 
manner still to be clarified) the set of all momentum measurements 
[the probability $w(p,t)$]. Interestingly, Torre~\cite{torre:randomness}, 
using a completely different approach, arrived at a similar conclusion, 
namely that the quantum mechanical 'variables' position and momentum 
cannot be random variables in the conventional sense. For simplicity 
we will continue to use the term random variable for $p$, and will 
add the attributes "`standard"' or ``nonstandard'' if required. 

As a first step in our study of $w(p,t)$, we will now investigate 
the integral equation~(\ref{eq:FIRSTAET}) and will derive a relation 
for  $w(p,t)$ which will be used again in 
section~\ref{sec:energyconservation}. In the course 
of the following calculations the behavior of $\rho$ and $S$ at 
infinity will frequently be required. We know that $\rho(x,t)$ is 
normalizable and vanishes at infinity. More specifically, we shall 
assume that $\rho(x,t)$ and $S(x,t)$ obey the following conditions:
\begin{equation}
  \label{eq:DERAGB3T}
 \rho\,A \to 0\mbox{,}\;\;
\frac{\partial \rho}{\partial x}\,A \to 0\mbox{,}\;\;
\frac{1}{\rho}\frac{\partial \rho}{\partial x}
\frac{\partial \rho}{\partial t} \to 0\mbox{,}\;\;
\;\;\mbox{for}\;x\to  ± \infty 
\mbox{,}
\end{equation}
where $A$ is anyone of the following factors
\begin{equation}
  \label{eq:AOOTFF}
1,\hspace{0.3cm}V,\hspace{0.3cm}\frac{\partial S}{\partial t}
,\hspace{0.3cm}x\frac{\partial S}{\partial x}
,\hspace{0.3cm}\left(\frac{\partial S}{\partial x}\right)^{2}
\mbox{.}
\end{equation}
Roughly speaking, condition~(\ref{eq:DERAGB3T}) means that 
$\rho$ vanishes faster than $1/x$ and $S$ is nonsingular at infinity. 
Whenever in the following  an integration by parts will be 
performed, one of the conditions~(\ref{eq:DERAGB3T}) will be used to 
eliminate the resulting boundary term. For brevity we shall not refer 
to~(\ref{eq:DERAGB3T}) any more; it will be sufficiently clear in the
context of the calculation which one of the factors in~(\ref{eq:AOOTFF})
will be referred to.  

We look for differential equations for our fields $\rho,\,S$ which 
are compatible with~(\ref{eq:FIRSTAET})-(\ref{eq:CONTONEDIM}).
According to the above discussion we are not allowed to 
identify~\eqref{eq:DEFMOMD1} with the random variable $p$. 
Using~(\ref{eq:CONTONEDIM}) we replace the derivative with respect 
to $t$ in~(\ref{eq:FIRSTAET}) by a derivative with respect to $x$ 
and perform an integration by parts. Then,~(\ref{eq:FIRSTAET}) takes 
the form     
\begin{equation}
  \label{eq:DFDCTNA}
 \int_{-\infty}^{\infty} \mathrm{d} x \rho(x,t) \frac{\partial S(x,t)}{\partial x} = \overline{p}
\mbox{.}
\end{equation}
Eq.~(\ref{eq:DFDCTNA}) shows that the averaged value of the 
random variable $p$ is the expectation value of the field $p(x,\,t)$. 
In the next section we shall insert this expression for $\overline{p}$ 
in the second statistical condition~\eqref{eq:SECONAET}. More specific 
results for the probability density $w(p,t)$ will be obtained later (in 
section~\ref{sec:derivationofL}). As an intermediate step, we now 
use~\eqref{eq:DFDCTNA} and~\eqref{eq:ERWAETP} to derive a relation 
for $w(p,t)$, introducing thereby an important change of variables. 

We replace the variables $\rho,\,S$ by new variables $\psi_1,\,\psi_2$ defined by
\begin{equation}
\psi_1=\sqrt{\rho}\,\cos \frac{S}{s}\hspace{0.5cm}
\psi_2=\sqrt{\rho}\,\sin \frac{S}{s}
\label{eq:FRSTHNZR}
\mbox{.}
\end{equation}
We may as well introduce the imaginary unit and define the complex 
field $\psi=\psi_1+\imath \psi_2$. Then, the last transformation and its inverse 
may be written as 
\begin{eqnarray}
\psi&=& \sqrt{\rho}\mathrm{e}^{\imath \frac{S}{s}}
\label{eq:FRSHAN3}\\
\rho=\psi \psi^{\star},&\hspace{0.3cm}&S = \frac{s}{2\imath}\ln \frac{\psi}{\psi^{\star}}
\label{eq:SCODAN3}
\mbox{.}
\end{eqnarray}
We note that so far no new condition or constraint has been introduced;  
choosing one of the sets of real variables $\{\rho,S\}$, $\{\psi_1,\psi_2\}$, or 
the set $\{\psi,\psi^{\star}\}$ of complex fields is just a matter of mathematical 
convenience. Using $\{\psi,\psi^{\star}\}$ the integrand on the left hand side 
of~(\ref{eq:DFDCTNA}) takes the form
\begin{equation}
  \label{eq:ZRL7NAB}
\rho\,\frac{\partial S}{\partial x}=\psi^{\star} \frac{s}{\imath}\frac{\partial}{\partial x}\psi-
\frac{s}{2\imath}\frac{\partial}{\partial x}|\psi|^{2}
\mbox{.}
\end{equation}
The derivative of $|\psi|^{2}$ may be omitted under the 
integral sign and~(\ref{eq:DFDCTNA}) takes the form 
\begin{equation}
  \label{eq:DGLFDPW}
\int_{-\infty}^{\infty} \mathrm{d} x \psi^{\star} \frac{s}{\imath}\frac{\partial}{\partial x}\psi = 
 \int_{-\infty}^{\infty} \mathrm{d} p w(p,t) p 
\mbox{.}
\end{equation}   
We introduce the Fourier transform of $\psi$, defined by  
\begin{eqnarray}
\psi(x,t)& = & \frac{1}{\sqrt{2\pi}a}
\int_{-\infty}^{\infty} \mathrm{d} \bar{p}\, \phi(\bar{p},t)\mathrm{e}^{\frac{\imath}{s} \bar{p} x}
\label{eq:DEFT5HIN}\\
\phi(\bar{p},t)& = & \frac{1}{\sqrt{2\pi}}
\int_{-\infty}^{\infty} \mathrm{d} x\, \psi(x,t)\mathrm{e}^{-\frac{\imath}{s} \bar{p} x}
\label{eq:DEFT5ZUR}
\mbox{.}
\end{eqnarray}
The constant $s$, introduced in Eq.~\eqref{eq:FRSTHNZR}, has the 
dimension of an action, which means that $\bar{p}$ has the dimension of 
a momentum. Performing the Fourier transform one finds that the  
momentum probability density may be written as
\begin{equation}
  \label{eq:ELMFU3GF}
w(p,t) = \frac{1}{s} |\phi(p,t)|^{2}+ h(p,t)
\mbox{,}
\end{equation}
where the integral over $ph(p,t)$ has to vanish. Using Parseval's 
formula and the fact that both $\rho(x,t)$ and $w(p,t)$ are normalized 
to unity we find that the integral of $h(p,t)$ has to vanish too.

Using the continuity equation~(\ref{eq:CONTONEDIM}) and the first 
statistical condition~(\ref{eq:FIRSTAET}) we found two results 
[namely~(\ref{eq:DGLFDPW}) and~\eqref{eq:ELMFU3GF}] which 
reduce for $h(p,t)=0$ to characteristic relations of the 
quantum mechanical formalism. However, the function $h(p,t)$, 
as well as the probability density $w(p,t)$ we are finally 
interested in, is still unknown, because the validity of the 
deterministic relation~\eqref{eq:DEFMOMD1} is not guaranteed in 
the present general formalism allowing for type 3 theories. In 
the next section the implications of the second statistical condition 
will be studied without using $w(p,t)$. We shall come back to the 
problem of the determination of $w(p,t)$ in section~\ref{sec:disorder}.

\section{Statistical theories}
\label{sec:statisticaltheories}
We study now the implications of the second statistical 
condition~(\ref{eq:SECONAET}). Using the variables $\rho,\,S$ it 
takes the form   
\begin{equation}
  \label{eq:DZDF23SE}
\frac{\mathrm{d}}{\mathrm{d}t}
 \int_{-\infty}^{\infty} \mathrm{d} x \rho \frac{\partial S}{\partial x} = 
- \int_{-\infty}^{\infty} \mathrm{d} x \rho \frac{\partial V}{\partial x}
\mbox{,}
\end{equation}
if $\overline{p}$ is replaced by the integral on the l.h.s. 
of~(\ref{eq:DFDCTNA}). Making again use of~(\ref{eq:CONTONEDIM}), we 
replace in~(\ref{eq:DZDF23SE}) the derivative of $\rho$ with respect 
to $t$ by a derivative with respect to $x$. Then, after an integration 
by parts, the left hand side of~(\ref{eq:DZDF23SE}) takes the form
\begin{equation}
  \label{eq:DLS34UIX}
\begin{split}
&\frac{\mathrm{d}}{\mathrm{d}t} \int_{-\infty}^{\infty} \mathrm{d} x \,\rho \frac{\partial S}{\partial x} = \\
& \int_{-\infty}^{\infty} \mathrm{d} x\,\left[
-\frac{1}{2m}\frac{\partial\rho}{\partial x}\left(\frac{\partial S}{\partial x} \right)^{2}+
\rho\frac{\partial}{\partial x}\frac{\partial S}{\partial t}
 \right]
\mbox{.}
\end{split}
\end{equation}
Performing two more integrations by parts [a second one 
in~(\ref{eq:DLS34UIX}) substituting the term with the time-derivative 
of $S$, and a third one on the right hand side of~(\ref{eq:DZDF23SE})], 
condition~(\ref{eq:SECONAET}) takes the final form 
\begin{equation}
  \label{eq:DZSGN23F}
 \int_{-\infty}^{\infty} \mathrm{d} x\,\frac{\partial\rho}{\partial x}\,\left[
\frac{1}{2m}\left(\frac{\partial S}{\partial x} \right)^{2}+
\frac{\partial S}{\partial t} + V \right]=0
\mbox{.}
\end{equation}
Equation~(\ref{eq:DZSGN23F}) can be considered as an integral equation 
for the real function $L(x,t)$ defined by 
\begin{equation}
  \label{eq:DDDGFLNEU}
L(x,t)=\frac{\partial S}{\partial t}+\frac{1}{2m}\left( \frac{\partial S(x,t)}{\partial x}\right)^{2}
+V(x,t)
\mbox{.}
\end{equation}
Obviously,~(\ref{eq:DZSGN23F}) admits an infinite number of 
solutions for $L(x,t)$, which are given by  
\begin{equation}
  \label{eq:DOEW2JNEU}
\frac{\partial\rho(x,t)}{\partial x}L(x,t)=\frac{\partial Q}{\partial x}
\mbox{,}
\end{equation}
The function $Q(x,t)$ in~(\ref{eq:DOEW2JNEU}) has to vanish at 
$x\to \pm \infty$ but is otherwise completely arbitrary.    

Equation~(\ref{eq:DOEW2JNEU}), with fixed $Q$ and $L$ as defined by
~(\ref{eq:DDDGFLNEU}), is the second differential equation for our 
variables $S$ and $\rho$ we were looking for, and defines - together with 
the continuity equation~(\ref{eq:CONTONEDIM}) - a statistical theory.
The dynamic behavior is completely determined by these differential equations 
for $S$ and $\rho$. On the other hand, the dynamic equation - in the sense 
of an equation describing the time-dependence of observable 
quantities - is given by~(\ref{eq:FIRSTAET}) and~(\ref{eq:SECONAET}).

From the subset of functions $Q$ which do not depend explicitely on $x$ 
and $t$ we list the following three possibilities for $Q$ and the 
corresponding $L$. The simplest solution is
\begin{equation}
  \label{eq:DELDICL}
Q=0,\hspace{2cm}L=0
\mbox{.}
\end{equation}
The second $Q$ depends only on $\rho$,
\begin{equation}
  \label{eq:DEMVTO7NEU}
Q \sim \rho^{n},\;\ n\geq{1},\;\;\;L \sim n\rho^{n-1}
\mbox{.}
\end{equation}
The third $Q$ depends also on the derivative of $\rho$,
\begin{equation}
  \label{eq:DZMFE8NEU}
Q \sim \frac{1}{2}\left(\frac{\partial}{\partial x}\sqrt{\rho} \right)^{2} ,\;\;\;
L \sim \frac{1}{2\sqrt{\rho}}\frac{\partial^{2}\sqrt{\rho}}{\partial x^{2}}
\mbox{.}
\end{equation}

We discuss first~(\ref{eq:DELDICL}). The statistical 
theory defined by~(\ref{eq:DELDICL}) consists of the continuity 
equation~(\ref{eq:CONTONEDIM}) and [see~(\ref{eq:DDDGFLNEU})] the 
Hamilton-Jacobi equation,
\begin{equation}
  \label{eq:HJGF4LUU}
\frac{\partial S}{\partial t}+\frac{1}{2m}\left( \frac{\partial S(x,t)}{\partial x}\right)^{2}
+V(x,t)=0
\mbox{.}
\end{equation} 
The fact that one of these equations agrees with the Hamilton-Jacobi 
equation does not imply that this theory is a type 1 theory (making 
predictions about individual events). This is not the case; many 
misleading statements concerning this limit may be found in the 
literature. It is a statistical theory whose observables are statistical 
averages. However, Eq.~(\ref{eq:HJGF4LUU}) becomes a type 1 theory 
if it is considered \emph{separately} - and embedded in the theory of 
canonical transformations. The crucial point is that~(\ref{eq:HJGF4LUU}) 
does not contain $\rho$; otherwise it could not be considered separately. 
This separability - or equivalently the absence of $\rho$ 
in~(\ref{eq:HJGF4LUU}) - implies that this theory is a classical (type 2) 
statistical theory~\cite{rosen:classical_quantum}. The function $S$ may 
be interpreted as describing the individual behavior of particles in 
the given environment (potential $V$). Loosely speaking, the function
$S$ may be identified with the considered particle; recall that $S$ is the
function generating the canonical transformation to a trivial Hamiltonian.
The identity of the particles described by $S$ is not influenced by 
statistical correlations because there is no coupling to $\rho$ 
in~(\ref{eq:HJGF4LUU}). The classical theory defined 
by~(\ref{eq:CONTONEDIM}) and~(\ref{eq:HJGF4LUU}) may also 
be formulated in terms of the variables $\psi$ and $\psi^{\star}$ [
but not as a single equation containing only $\psi$; see the 
remark at the end of section~(\ref{sec:statisticaltheories})]. 
In this form it has been discussed in several 
works~\cite{schiller:quasiclassical,rosen:classical_quantum,nikolic:classical}. 

All theories with nontrivial $Q$, depending on $\rho$ or its derivatives, 
should be classified as ``non-classical'' (or type 3) according to the 
above analysis. In non-classical theories any treatment of single 
events (calculation of trajectories) is impossible due to the coupling 
between $S$ and $\rho$. The problem is that single events are nevertheless 
real and observable. There must be a kind of dependence (correlation 
of non-classical type) between these single events. But this dependence 
cannot be described by concepts of deterministic theories 
like ``interaction''.           

The impossibility to identify objects in type 3 theories - independently 
from the statistical context - is obviously related to the breakdown 
of the concept of standard random variables discussed in the last 
section. There, we anticipated that a standard random variable (which is 
defined as a unique function of another random variable) contains 
an element of determinism that should be absent in type 3 theories. 
In fact, it does not make sense to define a unique relation between 
measuring data - e.g. of spatial position and momentum - if the 
quantities to be measured cannot themselves be defined independently 
from statistical aspects. 

The theory defined by Eq.~(\ref{eq:DEMVTO7NEU}) is a type 3 theory. We will
not discuss it in detail because it may be shown (see the next section) to 
be unphysical. It has been listed here in order to have a concrete example 
from the large set of insignificant type 3 theories.
 
The theory defined by Eq.~(\ref{eq:DZMFE8NEU}) is also a type 3 theory. 
Here, the second statistical condition takes the form 
\begin{equation}
  \label{eq:SKID5HJQ}
\frac{\partial S}{\partial t} + \frac{1}{2m}
\left(\frac{\partial S}{\partial x}\right)^{2}+V-
\frac{\hbar^2}{2m} \frac{1}{\sqrt{\rho}} \frac{\partial^{2}\sqrt{\rho}}{\partial x^{2}} =0
\mbox{,}
\end{equation}
if the free proportionality constant in~(\ref{eq:DZMFE8NEU}) is 
fixed according to $\hbar^{2}/m$. The two equations~(\ref{eq:CONTONEDIM})
and~(\ref{eq:SKID5HJQ}) may be rewritten in a more familiar form if 
the transformation~(\ref{eq:SCODAN3}) (with $s=\hbar$) to variables 
$\psi,\,\psi^{\star}$ is performed. Then, both equations are contained 
(as real and imaginary parts) in the single equation 
\begin{equation}
  \label{eq:SCH43STRD}
-\frac{\hbar}{\imath} \frac{\partial\psi}{\partial t}=
-\frac{\hbar^{2}}{2m}\frac{\partial^{2}\psi}{\partial x^{2}}+V\psi
\mbox{,}
\end{equation}
which is the one-dimensional version of Schr{\"o}dinger's 
equation~\cite{schrodinger:quantisierung_I}. Thus, 
quantum mechanics belongs to the class of theories 
defined by the above conditions. We see that the 
statistical conditions~(\ref{eq:FIRSTAET}), (\ref{eq:SECONAET})
comprise both quantum mechanical and classical statistical theories; 
these relations express a ``deep-rooted unity''~\cite{sen_et_al:ehrenfest} 
of the classical and quantum mechanical domain of physics.

We found an infinite number of statistical theories which are all
compatible with our basic conditions and are all on equal footing
so far. However, only one of them, quantum mechanics, is realized by 
nature. This situation leads us to ask which further conditions are 
required to single out quantum mechanics from this set. Knowing such
condition(s) would allow us to have premises which imply quantum 
mechanics. 

The above analysis shows that Schr{\"o}dinger's equation~(\ref{eq:SCH43STRD}) 
can be derived from the condition that the dynamic law for the probabilities 
takes the form of a \emph{single} equation for $\psi$ (instead of two 
equations for $\psi$ and $\psi^{\star}$ as is the case for all other 
theories). Our previous use of the variables $\psi$ and $\psi^{\star}$ 
instead of $S$ and $\rho$ was entirely a matter of mathematical 
convenience. In contrast, this last condition presents a real 
constraint for the physics since a different number field has 
been chosen~\cite{khrennikov:interference}. Recently, Schr{\"o}dingers 
equation including the gauge coupling term has been 
derived~\cite{klein:schroedingers} from this condition (which had 
to be supplemented by two further conditions, namely the existence 
of a continuity equation and the assumption of a linear time evolution 
law for $\psi$). Of course, this is a mathematical condition whose physical 
meaning is not at all clear. This formal criterion will be replaced in 
section~\ref{sec:subsidiary-condition} by a different condition which 
leads to the same conclusion but may be formulated in more physical 
terms.
\section{Energy conservation}
\label{sec:energyconservation}
In the last section [section~(\ref{sec:statisticaltheories})] we derived
a second differential equation~(\ref{eq:DOEW2JNEU}) for our dynamical 
variables $\rho$ and $S$. This equation has some terms in common with 
the Hamilton-Jacobi equation of classical mechanics but contains an 
unknown function $Q$ depending on $\rho$ and $S$; in principle it could 
also depend on $x$ and $t$ but this would contradict the homogeneity of 
space-time. We need further physical condition(s) to determine those 
functions $Q$ which are appropriate for a description of quantum mechanical 
reality or its classical counterpart. 

A rather obvious requirement is conservation of energy. In deterministic 
theories conservation laws - and in particular the energy conservation 
law which will be considered exclusively here - are a logical consequence 
of the basic equations; there is no need for separate postulates in this 
case. In statistical theories energy conservation with regard to 
time-dependence of single events is of course meaningless. However, a 
statistical analog of this conservation law may be formulated as 
follows: ``The statistical average of the random variable energy is 
time-independent''. In the present framework it is expressed by 
the relation 
\begin{equation}
  \label{eq:DFCONSENST}
\frac{\mathrm{d}}{\mathrm{d}t}
\left[
 \int_{-\infty}^{\infty} \mathrm{d} p w(p,t) \frac{p^{2}}{2m}+
 \int_{-\infty}^{\infty} \mathrm{d} x \rho(x,t) V(x)
 \right]=0
\mbox{.}
\end{equation}
We will use the abbreviation $\overline{E}=\overline{T}+\overline{V}$ 
for the bracket where $\overline{T}$ denotes the first and 
$\overline{V}$ denotes the second term respectively. Here, 
in contrast to the deterministic case, the fundamental laws 
[namely~(\ref{eq:FIRSTAET}),~(\ref{eq:SECONAET}),~(\ref{eq:CONTONEDIM})]
do not guarantee the validity of~(\ref{eq:DFCONSENST}). It has to be 
implemented as a separate statistical condition. In fact, 
Eq.~(\ref{eq:DFCONSENST}) is very simple and convincing; it seems 
reasonable to keep only those statistical theories which obey the 
statistical version of the fundamental energy conservation law. 

In writing down Eq.~\eqref{eq:DFCONSENST} a second tacit assumption,
besides the postulate of energy conservation, has been 
made, namely that a standard random variable ``kinetic energy'' exists; 
this assumption has already been formulated and partly justified in the 
last section. This means, in particular, that the probability density 
$w(p,t)$, which has been introduced in the statistical 
conditions~\eqref{eq:FIRSTAET},~(\ref{eq:SECONAET}) to obtain the 
expectation value of $p$ may also be used to calculate the expectation 
value of $p^{2}$. This second assumption is - like the requirement of 
energy conservation - not a consequence of the basic 
equations~\eqref{eq:DOEW2JNEU},~\eqref{eq:CONTONEDIM}. 
The latter may be used to calculate the probability density $\rho$ 
but says nothing about the calculation of expectation values of 
$p-$dependent quantities. Thus, Eq.~\eqref{eq:DFCONSENST} is an 
additional assumption, as may also be seen by the fact that two 
unknown functions, namely $h$ and $Q$ occur in~\eqref{eq:DFCONSENST}. 

Eq.~(\ref{eq:DFCONSENST}) defines a relation between $Q$ and $h$. More 
precisely, we consider variables $\rho$ and $S$ which are solutions of 
the two basic equations~(\ref{eq:DOEW2JNEU}) 
and ~(\ref{eq:CONTONEDIM}), where $Q$ may be an arbitrary function of $\rho$, 
$S$. Using these solutions we look which (differential) 
relations between $Q$ and $h$ are compatible with the 
requirement~(\ref{eq:DFCONSENST}). Postulating the validity 
of~(\ref{eq:DFCONSENST}) implies certain relations (yet 
to be found in explicit form) between the equations determining the 
probabilities and the equations defining the expectation values of 
$p$-dependent quantities (like the kinetic energy).

In a first step we rewrite the statistical average of $p^{2}$ 
in~(\ref{eq:DFCONSENST}) using~\eqref{eq:ELMFU3GF}. The result is
\begin{equation}
  \label{eq:AK34LFZD}
 \int_{-\infty}^{\infty} \mathrm{d} p w(p,t) p^{2}=
 \int_{-\infty}^{\infty} \mathrm{d} x \psi^{\star} 
\left( \frac{s}{\imath}\frac{\partial}{\partial x}\right)^{2}\psi 
+\int_{-\infty}^{\infty} \mathrm{d} p\, h(p,t) p^{2}
\mbox{,}
\end{equation}
as may be verified with the help of~(\ref{eq:DEFT5ZUR}).
Using~(\ref{eq:AK34LFZD}), transforming to $\rho,\,S$, and 
performing  an integration by parts, the first term 
of~(\ref{eq:DFCONSENST}) takes the form 
\begin{equation}
  \label{eq:AZWG4ZZUA}
\begin{split}
&\frac{\mathrm{d}\overline{T}}{\mathrm{d}t}=
\frac{1}{2m} \int_{-\infty}^{\infty} \mathrm{d} x \,
\Bigg\{ 
\bigg[
\frac{s^{2}}{4\rho^{2}} \left(\frac{\partial\rho}{\partial x}\right)^{2} 
- \frac{s^{2}}{2\rho}  \frac{\partial^{2}\rho}{\partial x^{2}} 
+\left( \frac{\partial S}{\partial x}\right)^{2}
\bigg] \frac{\partial\rho}{\partial t}\\
&-2\bigg[ 
\frac{\partial\rho}{\partial x}\frac{\partial S}{\partial x}+ \rho\frac{\partial^{2} S}{\partial x^{2}}
\bigg]\frac{\partial S}{\partial t}
\Bigg\}
+ \int_{-\infty}^{\infty} \mathrm{d} p \, \frac{\partial h(p,t)}{\partial t} \frac{p^{2}}{2m}
\mbox{.}
\end{split}
\end{equation}
If we add the time derivative of $\overline{V}$ 
to~(\ref{eq:AZWG4ZZUA})  we obtain the time derivative of 
$\overline{E}$, as defined by the left hand side 
of~(\ref{eq:DFCONSENST}). In the integrand of the latter expression 
the following term occurs
\begin{equation}
  \label{eq:AGAJ7GGL}
\left[ \left( \frac{\partial S}{\partial x}\right)^{2} +2mV \right]\frac{\partial\rho}{\partial t}
-2\left[\frac{\partial\rho}{\partial x}\frac{\partial S}{\partial x}+ \rho\frac{\partial^{2} S}{\partial x^{2}} 
\right]\frac{\partial S}{\partial t} 
\mbox{.}
\end{equation}
The two brackets in~(\ref{eq:AGAJ7GGL}) may be rewritten with the 
help of~(\ref{eq:CONTONEDIM}) and~(\ref{eq:DOEW2JNEU}). Then, the 
term~(\ref{eq:AGAJ7GGL}) takes the much simpler form
\begin{equation}
  \label{eq:RDB4EGT}
2m\left( \frac{\partial\rho}{\partial x}\right)^{-1} \frac{\partial Q}{\partial x}\frac{\partial \rho}{\partial t}
\mbox{.}
\end{equation}
Using~(\ref{eq:AZWG4ZZUA}) and~(\ref{eq:RDB4EGT}) we find that 
the statistical condition~(\ref{eq:DFCONSENST}) implies the 
following integral relation between $Q$ and $h$.
\begin{equation}
  \label{eq:HIF2IFJ}
\begin{split}
&\int_{-\infty}^{\infty} \mathrm{d} x \,
\bigg[
\left( \frac{\partial\rho}{\partial x}\right)^{-1} \frac{\partial Q}{\partial x}-
\frac{s^2}{2m} \frac{1}{\sqrt{\rho}} \frac{\partial^{2}\sqrt{\rho}}{\partial x^{2}}
\bigg] \frac{\partial\rho}{\partial t}\\ 
&+ \int_{-\infty}^{\infty} \mathrm{d} p \, \frac{\partial h(p,t)}{\partial t} \frac{p^{2}}{2m}
=0 
\mbox{.}
\end{split}
\end{equation}

Let us first investigate the classical solution. We may either insert 
the classical, ``hybrid'' solution~\eqref{eq:asff8eq} for $w(p,t)$ 
directly into Eq.\eqref{eq:DFCONSENST} or insert $h(p,t)$ according 
to~\eqref{eq:ELMFU3GF} with $w(p,t)$ as given by~\eqref{eq:asff8eq} 
in~\eqref{eq:HIF2IFJ}, to obtain   
\begin{equation}
  \label{eq:ARSEFEC}
\int_{-\infty}^{\infty} \mathrm{d} x \,
\left( \frac{\partial\rho}{\partial x}\right)^{-1} \frac{\partial Q}{\partial x} \frac{\partial\rho}{\partial t}=0
\mbox{,}
\end{equation}
which implies $\partial Q/ \partial x =0$. Thus, the hybrid probability 
density~\eqref{eq:asff8eq}  leads, as expected, to a classical (the 
equation for $S$ does not contain terms dependent on $\rho$) statistical 
theory, given by the Hamilton-Jacobi equation and the continuity 
equation. These equations constitute the classical limit of quantum 
mechanics which is a statistical theory (of type 2 according to the 
above classification) and not a deterministic (type 1) theory like 
classical mechanics. This difference is very important and should be 
borne in mind. The various 
ambiguities~\cite{kovner.rosenstein:quantisation} one encounters in 
the conventional particle picture both in the transitions from classical 
physics to quantum mechanics and back to classical physics, do not exist 
in the present approach. 

If we insert the quantum-mechanical result~\eqref{eq:DZMFE8NEU} with 
properly adjusted constant in~\eqref{eq:HIF2IFJ}, we obtain
\begin{equation}
  \label{eq:JJ3RTES}
 \int_{-\infty}^{\infty} \mathrm{d} p \, h(p,t) \frac{p^{2}}{2m}=T_{0}
\mbox{,}
\end{equation}
where $T_{0}$ is an arbitrary time-independent constant. This constant 
reflects the possibility to fix a zero point of a (kinetic energy) scale.
An analogous arbitrary constant $V_{0}$ occurs for the potential energy.
Since kinetic energy occurs always (in all physically meaningful contexts)
together with potential energy, the constant $T_{0}$ may be eliminated 
with the help of a properly adjusted $V_{0}$. Therefore, we see that -  as 
far as the calculation of the expectation value of the kinetic energy is 
concerned - it is allowed to set $h=0$. Combined with previous results, we 
see that $h$ may be set equal to $0$ as far as the calculation of the 
expectation values of $p^{n}$, for $n=0,\,1,\,2$ is concerned. These 
cases include all cases of practical importance. A universal rule for 
the calculation of averages of arbitrary powers of $p$ is not available in 
the present theory. The same is true for arbitrary powers of  $x$ and $p$.
Fortunately, this is not really a problem since the above powers cover all 
cases of physical interest, as far as powers of $p$ are concerned 
(combinations of powers of $x$ and $p$ do not occur in the present 
theory and will be dealt with in a future work). 

It is informative to compare the present theory with the corresponding 
situation in the established formulations of quantum mechanics. In the 
conventional quantization procedure, which is ideologically dominated 
by the structure of particle mechanics, it is postulated that all 
classical observables (arbitrary functions of $x$ and $p$) be represented 
by operators in Hilbert space. The explicit construction of these operators 
runs into considerable difficulties~\cite{shewell:formation} for all except 
the simplest combinations of $x$ and $p$. But, typically, this does not 
cause any real problems since all simple combinations (of physical interest) 
can be represented in a unique way by corresponding operators. Thus, what 
is wrong - or rather ill-posed - is obviously the postulate itself, which 
creates an artificial problem. This is one example, among several others, 
for an artificial problem created by choosing the  wrong (deterministic) 
starting point for quantization.  

If we start from the r.h.s. of~(\ref{eq:HIF2IFJ}) and postulate $h=0$, 
then we obtain agreement with the standard formalism of quantum mechanics, 
both with regard to the time evolution equation and the rules for 
calculating expectation values of $p-$dependent quantities. Thus, $h=0$ is 
a rather strong condition. Unfortunately, there seems be no intuitive 
interpretation at all for this condition. It is even less understandable than 
our previous formal postulate leading to Schr{\"o}dinger's equation, the 
requirement of a complex state variable. Thus, while we gained in this 
section important insight in the relation between energy conservation, 
time-evolution equation and rules for calculating expectation values, 
still other methods are required if we want to derive quantum mechanics 
from a set of physically interpretable postulates.

\section{Entropy as a measure of disorder ?}
\label{sec:disorder}

How then to determine the unknown function $h(p,t)$ [and $w(p,t)$] ? 
According to the last section, all the required information on 
$h(p,t)$  may be obtained from a knowledge of the term $Q$ in the 
differential equation~(\ref{eq:DOEW2JNEU}) for $\rho(x,t)$. We shall 
try to solve this problem my means of the following two-step strategy:
(i) Find an additional physical condition for the fundamental probability 
density $\rho(x,t)$, (ii) determine the shape of $Q$ [as well as that 
of $h(p,t)$ and $w(p,t)$] from this condition.

At this point it may be useful to recall the way probability 
densities are determined in classical statistical physics. 
After all, the present class of theories is certainly not of a 
deterministic nature and belongs fundamentally to the same class of 
statistical (i.e. incomplete with regard to the description of single 
events) theories as classical statistical physics; no matter how 
important the remaining differences may be. 

The physical condition for $\rho$ which determines the behavior of 
ensembles in classical statistical physics is the principle of 
\emph{maximal} (Boltzmann) \emph{entropy}. It agrees essentially with 
the information-theoretic measure of disorder introduced by 
Shannon~\cite{shannon:communication}. Using this principle both 
the micro-canonical and the canonical distribution of statistical 
thermodynamics may be derived under appropriate constraints.
Let us discuss this classical extremal principle in some detail in order 
to see if it can be applied, after appropriate modifications, to the 
present problem. This question also entails a comparison of different 
types of statistical theories. 

The Boltzmann-Shannon entropy is defined as a functional $S[\rho]$ of 
an arbitrary probability density $\rho$. The statistical properties 
characterizing disorder, which may be used to define this functional, 
are discussed in many 
publications~\cite{ben-naim:entropy},~\cite{khinchin:mathematical}. 
Only one of these conditions will, for later use, be written down 
here, namely the so-called ``composition law'': Let us assume 
that $\rho$ may be written in the form $\rho = \rho_1 \rho_2 $ where 
$\rho_i,\;i=1,2$ depends only on points in a subspace $X_i$ of our 
$n-$dimensional sample space $X$ and let us further assume that $X$ 
is the direct product of $X_1$ and $X_2$. Thus, this system consists 
of two independent subsystems. Then, the composition law is given by
\begin{equation}
  \label{eq:ASCLIAA}
S[\rho_1 \rho_2]=S^{(1)}[\rho_1] + S^{(2)}[\rho_2]
\mbox{,}
\end{equation}
where $S^{(i)}$ operates only on $X_i$.

For a countable sample space with events labeled by indices $i$ from 
an index set $I$ and probabilities $\rho_{i}$, the entropy is given by 
\begin{equation}
  \label{eq:DSDFOE4}
S[\rho]=-k \sum_{i \in I}\,\rho_i \ln \rho_i
\mbox{,}
\end{equation}
where $k$ is a constant. To obtain meaningful results the 
extrema of~(\ref{eq:DSDFOE4}) under appropriate constraints, or subsidiary 
conditions, must be found. The simplest constraint is the normalization 
condition $\sum \rho_i =1$. In this case the extrema of the function 
\begin{equation}
  \label{eq:AWKF34FE}
F[\rho,\lambda]=-k \sum_{i \in I}\,\rho_i \ln \rho_i+
\lambda \left(\sum_{i \in I}\,\rho_i-1 \right)
\mbox{}
\end{equation}
with respect to the variables $\rho_{1},\,...\rho_{N},\,\lambda$ must be 
calculated. One obtains the reasonable result that the minimal value 
of $F[\rho,\lambda]$ is $0$ (one of the $\rho_{i}$ equal to $1$, all 
other equal to $0$) and the maximal value is $k\ln N$  (all $\rho_{i}$ 
equal, $\rho_{i}=1/N$).
 
For most problems of physical interest the sample space is non-denumerable. 
A straightforward generalization of Eq.~\eqref{eq:DSDFOE4} is given by 
\begin{equation}
  \label{eq:UJNDDOE2}
S[\rho]=-k\int \mathrm{d}x\,\rho(x) \ln \rho(x)
\mbox{,}
\end{equation}  
where the symbol $x$ denotes now a point in the appropriate (generally 
$n-$dimensional) sample space. There are some problems inherent in 
the this straightforward transition to a continuous set of events which 
will be mentioned briefly in the next section. Let us put aside this 
problems for the moment and ask if~\eqref{eq:UJNDDOE2} makes sense 
from a physical point of view. For non-denumerable problems the principle 
of maximal disorder leads to a variational problem and the method of 
Lagrange multipliers may still be used to combine the requirement of 
maximal entropy with other defining properties (constraints). An 
important constraint is the property of constant temperature which 
leads to the condition that the expectation value of the possible 
energy values $E(x)$ is given by a fixed number $\bar{E}$,
\begin{equation}
  \label{eq:TGDP72E}
\bar{E} = \int \mathrm{d}x\,\rho(x) E(x)
\mbox{,}
\end{equation}
If, in addition, normalizability is implemented as a defining property, then 
the true distribution should be an extremum of the functional  
\begin{equation}
  \label{eq:DCV88DSP}
K[\rho]=-k\int \mathrm{d}x\,\rho(x) \ln \rho(x)
-\lambda_{2} \int \mathrm{d}x\,\rho(x) E(x)
-\lambda_{1} \int \mathrm{d}x\,\rho(x)
\mbox{.}
\end{equation}
It is easy to see that the well-known canonical distribution of 
statistical physics is indeed an extremum of $K[\rho]$. 
Can we use a properly adapted version of this powerful principle 
of maximal disorder (entropy) to solve our present problem ?  

Let us compare the class of theories derived in 
section~\ref{sec:statisticaltheories} with classical theories
like~\eqref{eq:DCV88DSP}. This may be of interest also in view of a 
possible identification of 'typical quantum mechanical properties' of 
statistical theories. We introduce for clarity some notation, based
on properties of the sample space. Classical statistical physics 
theories like~\eqref{eq:DCV88DSP} will be referred to as ''phase 
space theories''. The class of statistical theories, derived in 
section~\ref{sec:statisticaltheories}, will be referred to as 
''configuration space theories''.  

The most fundamental difference between phase space theories and 
configuration space theories concerns the physical meaning of the 
coordinates. The coordinates $\mathbf{x}$ of phase space theories are  
(generally time-dependent) labels for particle properties. In contrast, 
configuration space theories are field theories; individual particles 
do not exist and the (in our case one-dimensional) coordinates $x$ are 
points in space. 

A second fundamental difference concerns the dimension of the sample 
space. Elementary events in phase space theories are points in phase 
space (of dimension $6$ for a $1-$particle system) including 
configuration-space \emph{and} momentum-space (particle) coordinates 
while the elementary events of configuration space theories are 
(space) points in configuration space (which would be of dimension $3$ 
for a $1-$ particle system in three spatial dimensions). This fundamental 
difference is a consequence of a (generally nonlocal) dependence  between 
momentum coordinates and space-time points contained in the postulates 
of the present theory, in particular in the postulated form of the probability 
current [see~\eqref{eq:CONTONEDIM}]. This assumption, a probability 
current, which takes the form of a gradient of a function $S$ (multiplied 
by $\rho$) is a key feature distinguishing configuration space theories, as 
potential quantum-like theories, from the familiar (many body) phase space
theories. The existence of this dependence per se is not an exclusive 
feature of quantum mechanics, it is a property of all theories belonging 
to the configuration class, including the theory characterized by 
$\hbar=0$, which will be referred to as ''classical limit theory''. What 
distinguishes the classical limit theory from quantum mechanics is the 
particular form of this dependence; for the former it is given by a 
conventional functional relationship (as discussed in 
section~\ref{sec:randomvariables}) for the latter it is given by 
a nonlocal relationship whose form is still to be determined. 

This dependence is responsible for the fact that no ''global'' condition 
[like~(\ref{eq:TGDP72E}) for the canonical distribution] must be introduced 
for the present theory in order to guarantee conservation 
of energy in the mean - this conservation law can be guaranteed ''locally'' 
for arbitrary theories of the configuration class by adjusting the 
relation between $Q$ (the form of the dynamic equation) and $h$ (the 
definition of expectation values). In phase space theories the form of 
the dynamical equations is fixed (given by the deterministic equations 
of classical mechanics). Under constraints like~(\ref{eq:TGDP72E}) the 
above principle of maximal disorder creates - basically by selecting 
appropriate initial conditions - those systems which belong to a 
particular energy; for non-stationary conditions the deterministic 
differential equations of classical mechanics guarantee then 
that energy conservation holds for all times. In contrast, in 
configuration space theories there are 
no initial conditions (for particles). The conditions which are at 
our disposal are the mathematical form of the expectation values 
(the function $h$) and/or the mathematical form of the differential 
equation (the function $Q$). Thus, if something like the principle of 
maximal disorder can be used in the present theory it will determine 
the form of the differential equation for $\rho$ rather than the explicit 
form of $\rho$. 

These considerations raise some doubt as to the usefulness of an 
measure of disorder like the entropy~\eqref{eq:UJNDDOE2} - which 
depends essentially on $E$ instead of $x$ and does not contain 
derivatives of $\rho$ - for the present problem. We may still look for 
an information theoretic extremal principle of the general form
\begin{equation}
  \label{eq:ZUG34EPR}
I[\rho]+\sum_{l} \lambda_lC_l[\rho]\to \mbox{extremum}
\mbox{.}
\end{equation}
Here, the functional $I[\rho]$ attains its maximal value for the 
function $\rho$ which describes - under given constraints $C_l[\rho]$ - the 
maximal disorder. But $I[\rho]$ will differ from the entropy functional 
and appropriate constraints $C_l[\rho]$, reflecting the local character 
of the present problem, have still to be found. Both terms 
in~\eqref{eq:ZUG34EPR} are at our disposal and will be defined 
in the next sections.     

\section{Fisher's information}
A second measure of disorder, besides entropy, exists which 
is called Fisher information~\cite{fisher:statistical*1}. The 
importance of this second type of ``entropy''for the mathematical 
form of the laws of physics - in particular for the terms related to 
the kinetic energy - has been stressed in a number of publications by 
Frieden and coworkers~\cite{frieden:fisher_basis,frieden:sciencefisher} 
and has been studied further by Hall~\cite{hall:quantumfisher}, 
Reginatto~\cite{reginatto:derivation} and others. The Fisher 
functional $I[\rho]$ is defined by 
\begin{equation}
  \label{eq:fisherfun}
I[\rho]=\int\mathrm{d}x\,\rho(x)\left(\frac{\rho^{\prime}(x)}{\rho(x)} \right)^{2}
\mbox{,}
\end{equation}
where $\rho^{\prime}$ denotes $\partial \rho  / \partial x$ in the present one-dimensional case,
and the $n-$component vector $\mathrm{D} \rho = ( \partial \rho  / \partial x_1,\ldots,\partial \rho  / \partial x_n )$ 
if $x=(x_1,\ldots,x_n)$. Since the time variable $t$ does not play an 
important role it will frequently be suppressed in this section.

The Boltzmann-Shannon entropy~(\ref{eq:UJNDDOE2}) and the Fisher 
information~(\ref{eq:fisherfun}) have a number of crucial 
statistical properties in common. We mention here, for 
future reference, only the most important one, namely the 
composition law~\eqref{eq:ASCLIAA}; a more complete list of common 
properties may be found in the literature~\cite{van:information}). 
Using the notation introduced in section~\ref{sec:disorder} [see 
the text preceeding Eq.~\eqref{eq:ASCLIAA}] it is easy to see 
that Eq.~\eqref{eq:fisherfun} fulfills the relation
\begin{equation}
  \label{eq:ASJU34AA}
I[\rho_1 \rho_2]=I^{(1)}[\rho_1] + I^{(2)}[\rho_2]
\mbox{,}
\end{equation}
in analogy to Eq.~(\ref{eq:ASCLIAA}) for the entropy $S$. The most 
obvious difference between~(\ref{eq:UJNDDOE2}) and~(\ref{eq:fisherfun}) 
is the fact that~(\ref{eq:fisherfun}) contains a derivative 
while~(\ref{eq:UJNDDOE2}) does not. As a consequence, extremizing 
these two functionals yields fundamentally different equations 
for $\rho$, namely a differential equation for the Fisher functional 
$I$ and an algebraic equation for the entropy functional $S$. 

The two measures of disorder, $S$ and $I$,  are related to each 
other. To find this relation, it is necessary to introduce 
a generalized version of~(\ref{eq:UJNDDOE2}), the so-called 
``relative entropy''. It is defined by
\begin{equation}
  \label{eq:KULLEIBE}
G[\rho,\alpha]=-\int \mathrm{d}x\,\rho(x) \ln \frac{\rho(x)}{\alpha(x)} 
\mbox{,}
\end{equation}
where $\alpha(x)$ is a given probability density, sometimes referred to 
as the ``prior'' [the constant $k$ in Eq.~(\ref{eq:UJNDDOE2}) has been 
suppressed here]. It provides a reference point for the unknown $\rho$; 
the best choice for $\rho$ is to be determined from the requirement of 
maximal relative entropy $G[\rho,\alpha]$ under given constraints, 
where $\alpha$ represents the state of affairs (or of our knowledge of 
the state of affairs) \emph{prior} to consideration of the constraints. 
The quantity $-G[\rho,\alpha]$ agrees with the  ``Kullback-Leibler distance'' 
between two probability densities  $\rho$ and 
$\alpha$~\cite{kullback:information}.

It has been pointed out that ``all entropies are relative 
entropies''~\cite{caticha:relative}. In fact, all physical quantities 
need reference points in order to become observables. The 
Boltzmann-Shannon entropy~(\ref{eq:UJNDDOE2}) is no exception. In this 
case, the `probability density' $\alpha(x)$ is a number of value $1$, 
and of the same dimension as $\rho(x)$; it describes absence of any knowledge 
or a completely disordered state. We mention also two other more technical 
points which imply the need for relative entropies. The first is the 
requirement to perform invariant variable transformations in the sample 
space~\cite{caticha:relative}, the second is the requirement to 
perform a smooth transition from discrete to continuous 
probabilities~\cite{jaynes:probability}. 

Thus, the concept of relative entropies is satisfying from a theoretical 
point of view. On the other hand it seems to be useless from a practical 
point of view since it requires - except in the trivial limit $\alpha=1$ - 
knowledge of a new function $\alpha(x)$ which is in general just as 
unknown as the original unknown function $\rho(x)$. A way out of this 
dilemma is to identify $\alpha(x)$ with a function $\rho(Tx)$, which can be 
obtained from $\rho(x)$ by replacing the argument $x$  by a transformed 
argument $Tx$. In this way we obtain from~(\ref{eq:KULLEIBE}) a quantity 
$G[\rho;p_T]$ which is a functional of the relevant function $\rho$ 
alone; in addition it is an ordinary function of the parameters 
$p_T$ characterizing the transformation. The physical meaning of the 
relative entropy remains unchanged, the requirement of maximal 
relative entropy $G[\rho;p_T]$ becomes a condition for the variation 
of $\rho$ in the sample space between the points $Tx$ and $x$.    

If further consideration is restricted to translations $Tx = x+\Delta x$ 
(it would be interesting to investigate other transformations, in 
particular if the sample space agrees with the configuration space)
then the relative entropy is written as 
\begin{equation}
  \label{eq:KLRELSPEC}
G[\rho;\Delta x] = 
-\int \mathrm{d}x\,\rho(x) \ln \frac{\rho(x)}{\rho(x+\Delta x)} 
\mbox{.}
\end{equation}
Expanding the integrand on the r.h.s. of~(\ref{eq:KLRELSPEC}) up to 
terms of second order in $\Delta x$ and using the fact that $\rho$ and 
$\rho^{\prime}$ have to vanish at infinity one obtains the relation  
\begin{equation}
  \label{eq:EHDRZFUE}
G[\rho;\Delta x] \doteq  - \frac{\Delta x^{2}}{2} I[\rho]
\mbox{.}
\end{equation}
This, then is the required relation between the relative entropy $G$ and 
and the Fisher information $I$; it is valid only for sufficiently small 
$\Delta x$. The relative entropy $G[\rho;\Delta x]$ cannot be positive. 
Considered as a function of $\Delta x$ it has a maximum at $\Delta x=0$ 
(taking its maximal value $0$) provided $I>0$. This means that the 
principle of maximal entropy implies \emph{no change at all} relative 
to an arbitrary reference density. This provides no criterion for 
$\rho$ since it holds for arbitrary $\rho$. But if~(\ref{eq:EHDRZFUE}) 
is considered, for fixed $\Delta x$, as a functional of $\rho$, the 
principle of maximal entropy implies, as a criterion for the spatial 
variation of $\rho$, a principle of \emph{minimal Fisher information}.
  
Thus, from this overview (see Frieden`s book~\cite{frieden:sciencefisher} 
for more details and several other interesting aspects) we would 
conclude that the 
principle of minimal Fisher information should not be considered as a 
completely new and exotic matter. Rather it should be considered as an 
extension or generalization of the classical principle of maximal 
disorder to a situation where a spatially varying probability exists, 
which contributes to disorder. This requires, in particular, 
that this probability density is to be determined from a differential 
equation and not from an algebraic equation. We conclude that the principle 
of minimal Fisher $I$ is very well suited for our present purpose. As a next 
step we have to set up proper constraints for the extremal principle.

\section{Subsidiary condition}
\label{sec:subsidiary-condition}
It will be convenient in the course of the following calculations 
to write the differential equation~(\ref{eq:DOEW2JNEU}) in the form 
\begin{equation}
  \label{eq:DVGDG8U}
L(x,t)-L_{0}(\rho,\rho^{\prime},\rho^{\prime \prime})=0
\mbox{,}
\end{equation}
where $L(x,t)$ is given by~(\ref{eq:DDDGFLNEU}) and $L_{0}$ is 
defined by  
\begin{equation}
  \label{eq:AESWFLNLL}
L_{0}(\rho,\rho^{\prime},\rho^{\prime \prime})=
\frac{Q^{\prime}(\rho,\rho^{\prime},\rho^{\prime \prime})}{\rho^{\prime}}
\mbox{.}
\end{equation}
In~(\ref{eq:AESWFLNLL}) it has been assumed that $L_{0}$ does not 
depend explicitely on $x,t$, that the problem is basically of a 
time-independent nature, and that no higher derivatives with respect 
to $x$ than $\rho^{\prime \prime}$ occur. This last assumption is in 
agreement with the mathematical form of all fundamental differential  
equations of physics; we shall come back to this point later. Our 
task is to determine the functional form of $L_{0}$, with respect 
to the variables $\rho,\,\rho^{\prime},\,\rho^{\prime \prime}$, using 
a general statistical extremal principle. As a consequence of the 
general nature of this problem we do not expect the solution to 
depend on the particular form of $L(x,t)$. For the same reason 
$L_{0}$ does not depend on $S$.    

We tentatively formulate a principle of maximal 
disorder of the form~(\ref{eq:ZUG34EPR}) and identify $I[\rho]$ with the
Fisher functional~(\ref{eq:fisherfun}). Then, the next step is 
to find a proper constraint  $C[\rho]$. In accord with 
general statistical principles the prescribed quantity should
have the form of a statistical average. A second condition 
is that our final choice should be as similar to the classical 
requirement~\eqref{eq:TGDP72E} as possible. Adopting these 
criteria one is led more or less automatically to the constraint  
\begin{equation}
  \label{eq:EAEC34MU}
C[\rho]=0
\mbox{,}
\end{equation}
where 
\begin{equation}
  \label{eq:ANKDRDDC}
C[\rho]=\int \mathrm{d}x\,\rho 
\left[ L(x,t)-L_{0}(\rho,\rho^{\prime},\rho^{\prime \prime}) \right]
\mbox{.}
\end{equation}
Our guideline in setting up this criterion has been the idea of a 
prescribed value of the average energy; the new term $L_0$ plays 
the role of an additional contribution to the energy. For 
$S(x,t)=-Et+S_0(x)$ and $L_0=0$ the constraint~(\ref{eq:EAEC34MU}) 
agrees with~(\ref{eq:TGDP72E}) provided the `classical` identification 
of $p$ with the gradient of $S_0(x)$ is performed [see~(\ref{eq:DEFMOMD1})]. 
The most striking difference between~(\ref{eq:TGDP72E}) 
and~(\ref{eq:EAEC34MU}) is the fact that the quantity $L-L_{0}$ 
in~(\ref{eq:ANKDRDDC}), whose expectation value yields 
the constraint, is not defined independently from the 
statistics [like $E$ in~(\ref{eq:TGDP72E})] but depends itself 
on $\rho$ (and its derivatives up to second order). This aspect
of non-classical theories has already been discussed in 
section~\ref{sec:statisticaltheories}.

Let us try to apply the mathematical apparatus of variational 
calculus~\cite{brunt:variations} to the constraint 
problem~(\ref{eq:ZUG34EPR}) with the "`entropy"' functional 
$I[\rho]$ defined by~(\ref{eq:fisherfun}) and a single 
constraint defined by~(\ref{eq:EAEC34MU}) and~(\ref{eq:ANKDRDDC})
(there is no normalization condition here because we do not want 
to exclude potentially meaningful non-normalizable states from 
the consideration). Here we encounter immediately a first problem 
which is due to the fact that our problem consists in the 
determination of an unknown function $L_0$ of $\rho,\rho^{\prime},\rho^{\prime \prime}$. 
This function appears in the differential equation \emph{and} in the 
subsidiary condition for the variational problem. Thus, our task is 
to identify from a variational problem the functional form of a 
constraint defining this variational problem. Variational calculus, 
starts, of course, from constraints whose functional forms
are \emph{fixed}; these fixed functionals are used to derive 
differential equations for the variable $\rho$. Thus, whenever 
the calculus of variations is applied, the function 
$L_{0}(\rho,\rho^{\prime},\rho^{\prime \prime})$ must be considered 
as \emph{unknown but fixed}. We shall have to find a way to 
'transform' the condition for the variation of $\rho(x)$ 
in a corresponding condition for the variation of 
$L_{0}(\rho,\rho^{\prime},\rho^{\prime \prime})$.   

The variational calculation defined above belongs to a class of 
'isoperimetric' variational problems which can be solved using 
the standard method of Lagrange multipliers, provided certain 
mathematical conditions are fulfilled~\cite{brunt:variations}. 
Analyzing the situation we encounter here a second problem, which 
is in fact related to the first. Let us briefly recall the way 
the variational problem~\eqref{eq:ZUG34EPR} is solved, in 
particular with regard to the role of the Lagrange multipliers 
$\lambda_i$~\cite{brunt:variations}. Given the problem to find an 
extremal $\rho_0(x)$ of $I[\rho]$ under $m$ constraints of the form 
$C_i[\rho]=0$ and two prescribed values of $\rho$ at the boundaries, 
one proceeds as follows. The Euler-Lagrange equation  belonging 
to the functional~\eqref{eq:ZUG34EPR} is solved. The general 
solution for $\rho$ depends (besides on $x$) on two integration 
constants, say $C_1$ and $C_2$, and on the $m$ Lagrange multipliers 
$\lambda_1,...,\lambda_m$. To obtain the final extremal $\rho_0(x)$, these 
$m+2$ constants have to be determined from the two boundary values and 
the $m$ constraints (which are differential equations for isoperimetric 
problems). This is exactly the way the calculation has been performed (even 
though a simpler form of the constraints has been used) in the classical 
case. For the present problem, however, this procedure is useless, since we 
do not want the constraints to determine the shape of individual solutions 
but rather the functional form of a term in the differential equation, 
which is then the same for all solutions. For that reason the `normal` 
variational problem~\eqref{eq:ZUG34EPR} does not work (we shall come back 
to a mathematical definition of `normal` and `abnormal` shortly). This means 
that the classical principle of maximal entropy, as discussed in 
section~\ref{sec:disorder}, cannot be taken over literally to 
the non-classical domain.

For the same reason, no subsidiary conditions can be taken into account 
in the calculations reported by Frieden~\cite{frieden:fisher_basis} and 
by Reginatto~\cite{reginatto:derivation}. In these works, a different route 
is chosen to obtain Schr{\"o}dingers equation; in contrast to the present 
work (see below) the Fisher functional is added as a new term to a classical 
Lagrangian and the particular form of this new term is justified by 
introducing a new "`principle of extreme physical 
information"'~\cite{frieden_soffer:lagrangians}.

A variational problem is called `normal` if an extremal of the functional
$I[\rho]+\lambda_1C[\rho]$ (here we restrict ourselves to the present 
case of a single constraint) exists which is \emph{not} at the same time an 
extremal of the constraint functional $C[\rho]$. If this is not the case, 
i.e. if the extremal \emph{is} at the same time an extremal of $C[\rho]$, 
then the problem is called `abnormal`~\cite{brunt:variations}. Then, the
usual derivation becomes invalid and the condition~(\ref{eq:ZUG34EPR}) 
must be replaced by the condition of extremal $C[\rho]$ alone,
\begin{equation}
  \label{eq:ZJU44EPR}
\lambda_1C[\rho]\to \mbox{extremum}
\mbox{,}
\end{equation}
which then yields  $\delta C[\rho] = 0$ as only remaining condition to determine 
the extremal. This type of problem is also sometimes referred to 
as ''rigid''; the original formulation~(\ref{eq:ZUG34EPR}) may be 
extended to include the abnormal case by introducing a second Lagrange 
multiplier~\cite{brunt:variations}.

We conclude that our present problem should be treated as an abnormal 
variational problem since we thereby get rid of our main difficulty, namely 
the unwanted dependence of individual solutions on Lagrange multipliers 
[$\lambda_1$ drops actually out of Eq.~(\ref{eq:ZJU44EPR})]. A somewhat
dissatisfying (at first sight) feature of this approach is the fact that 
the Fisher functional $I[\rho]$ itself does no longer take part in the 
variational procedure; the original idea of implementing maximal 
statistical disorder seems to have been lost. But it turns out that we
shall soon recover the Fisher $I$ in the course of the following 
calculation. The vanishing of the first variation of $C[\rho]$, written 
explicitely as
\begin{equation}
  \label{eq:AJ8UZ4DC}
\delta C[\rho]=\delta \int \mathrm{d}x\,\rho 
\left[ L(x,t)-L_{0}(\rho,\rho^{\prime},\rho^{\prime \prime}) \right]=0
\mbox{,}
\end{equation}
means that (for fixed $L_0$) the spatial variation of $\rho$ should
extremize (minimize) the average value of the deviation from $L(x,t)$.
This requirement is [as a condition for $\rho(x)$] in agreement with 
the principle of minimal Fisher information as a special realization 
of the requirement of maximal disorder. Eq.~(\ref{eq:AJ8UZ4DC}) defines 
actually a Lagrangian for $\rho $ and yields as Euler-Lagrange equations a 
differential equation for $\rho $. When this equation is derived 
the task of variational calculus is finished. On the other hand, we 
know that $\rho$ obeys also Eq.~(\ref{eq:DVGDG8U}). Both differential 
equations must agree and this fact yields a condition for our unknown 
function $L_0$. Eq.~~(\ref{eq:DVGDG8U}) also guarantees that the 
original constraint~(\ref{eq:EAEC34MU}) is fulfilled. In this way 
we are able to 'transform' the original variational condition for 
$\rho(x)$ in a condition for $L_{0}(\rho,\rho^{\prime},\rho^{\prime \prime})$. 
In the next section this condition will be used to calculate  $L_0$ and 
to recover the form of the Fisher $I$.

It should be mentioned that Eq.~(\ref{eq:AJ8UZ4DC}) has been used many 
times in the last eighty years to derive Schr{\"o}dinger's equation from 
the Hamilton-Jacobi equation. The first and most important of these works 
is Schr{\"o}dinger's "`Erste Mitteilung"'~\cite{schrodinger:quantisierung_I}.
In all of these papers $L_0$ is not treated as an unknown function but as
a \emph{given} function, constructed with the help of the 
following procedure. 
First, a transformation from the variable $S$ to a complex variable 
$\psi^{\prime}=\mathrm{exp}[\imath S/\hbar] $ is performed. Secondly, a 
\emph{new} variable $\rho$ is introduced by means of the formal replacement 
$\psi^{\prime}\Rightarrow \psi=\rho\psi^{\prime}$. This creates a new term 
in the Lagrangian, which has exactly the form required to create 
quantum mechanics. More details 
on the physical motivations underlying this replacement procedure 
may be found in a paper by Lee and Zhu~\cite{lee.zhu:principle}.
It is interesting to note that the same formal replacement may be used 
to perform the transition from the London theory of superconductivity to 
the Ginzburg-Landau theory~\cite{klein:londongl}. There, the necessity 
to introduce a new variable is obvious, in contrast to the present much 
more intricate situation. 

\section{Derivation of $L_0$ }
\label{sec:derivationofL}

For a general $L_{0}$, the Euler-Lagrange equation belonging to 
the functional $C[\rho]$ [see~(\ref{eq:AJ8UZ4DC})] depends on 
derivatives higher than second order since the integrand 
in~(\ref{eq:AJ8UZ4DC}) depends on $\rho^{\prime \prime}$. This is 
a problem, since according to the universal rule mentioned above 
all differential equations of physics are formulated using 
derivatives not higher than second order. If we are to 
conform with this general rule 
(and we would like to do so) then we should use a Lagrangian 
containing only first order derivatives. But this would then
again produce a conflict with Eq.~(\ref{eq:DVGDG8U}) because the 
variational procedure increases the order of the highest 
derivative by one. We postpone the resolution of this 
conflict and proceed by calculating the Euler-Lagrange equations 
according to~(\ref{eq:AJ8UZ4DC}), which are given by
\begin{equation}
  \label{eq:DDWGUE56F}
\begin{split}
L(x,t)&-L_{0}(\rho,\rho^{\prime},\rho^{\prime \prime})\\ 
-\frac{\mathrm{d}^{2}}{\mathrm{d}x^{2}}
\rho \frac{\partial L_0}{\partial \rho^{\prime \prime}}
+\rho^{\prime} \frac{\partial L_0}{\partial \rho^{\prime}}
&+\rho \frac{\mathrm{d}}{\mathrm{d}x}
\frac{\partial L_0}{\partial \rho^{\prime}}
-\rho \frac{\partial L_0}{\partial \rho}=0
\mbox{.}
\end{split}
\end{equation}
Using the second basic condition~(\ref{eq:DVGDG8U}) we see that 
the first line of Eq.~(\ref{eq:DDWGUE56F}) vanishes and we obtain, 
introducing the abbreviation $\beta=\rho L_0$, the following partial 
differential equation for the determination of the functional 
form of $L_0$ with respect to the variables 
$\rho,\rho^{\prime},\rho^{\prime \prime}$,   
\begin{equation}
  \label{eq:DWLFF98DX}
-\frac{\mathrm{d}^{2}}{\mathrm{d}x^{2}}
\frac{\partial \beta}{\partial \rho^{\prime \prime} }
+\frac{\mathrm{d}}{\mathrm{d}x}
\frac{\partial \beta}{\partial \rho^{\prime}}
-\frac{\partial \beta}{\partial \rho}
+\frac{\beta}{\rho}=0
\mbox{.}
\end{equation}
Expressing the derivatives of $L_0$ in terms of the derivatives of
$\rho,\rho^{\prime}$ and $\rho^{\prime \prime}$ leads to a lengthy relation which will not be 
written down here. Since $L_0$ does not contain
higher derivatives than $\rho^{\prime \prime}$, the sums of the coefficients 
of both the third and fourth derivatives of $\rho$ have to vanish. This 
implies that $\beta$ may be written in the form
\begin{equation}
  \label{eq:IB23AMF7F}
\beta(\rho,\rho^{\prime},\rho^{\prime \prime})=
C(\rho,\rho^{\prime})\rho^{\prime \prime}+D(\rho,\rho^{\prime})
\mbox{,}
\end{equation}
where $C(\rho,\rho^{\prime})$ and  $D(\rho,\rho^{\prime})$ are 
solutions of 
\begin{eqnarray}
-2\frac{\partial C}{\partial \rho}
-\frac{\partial^{2}C}{\partial \rho \partial \rho^{\prime}} \rho^{\prime}
+\frac{\partial^{2}D}{\partial (\rho^{\prime})^{2}}
+\frac{C}{\rho} = &0& 
\label{eq:DEI78HMA}\\
-\frac{\partial^{2}C}{\partial \rho^{2}} (\rho^{\prime})^{2}
+\frac{\partial^{2}D}{\partial \rho \partial \rho^{\prime}}\rho^{\prime}
-\frac{\partial D}{\partial \rho}
+\frac{D}{\rho}& = &0 
\label{eq:DEO99HMA}
\mbox{.}
\end{eqnarray}
Thus, two functions of $\rho,\rho^{\prime}$ have to be found, instead
of a single function of $\rho,\rho^{\prime},\rho^{\prime \prime}$. 
Fortunately, the solution we look for presents a term in a differential
equation. This allows us to restrict our search to relatively simple
solutions of~\eqref{eq:DEI78HMA},~\eqref{eq:DEO99HMA}. If the differential 
equation is intended to be comparable in complexity to other fundamental 
laws of physics, then a polynomial form,
\begin{equation}
  \label{eq:AE12CI9C}
C(\rho,\rho^{\prime})=\sum_{n,m=-\infty}^{\infty} 
c_{n,m}\rho^{n} (\rho^{\prime})^{m},\;\;\;
D(\rho,\rho^{\prime})=\sum_{n,m=-\infty}^{\infty}
d_{n,m}\rho^{n} (\rho^{\prime})^{m}
\mbox{,}
\end{equation}
preferably with a finite number of terms, will be a sufficiently
general Ansatz.

Eqs.~\eqref{eq:DEI78HMA} and \eqref{eq:DEO99HMA} must of course hold 
for arbitrary $\rho,\rho^{\prime}$. Inserting the Ansatz~\eqref{eq:AE12CI9C}, 
renaming indices and comparing coefficients of equal powers of $\rho$ and 
$\rho^{\prime}$ one obtains the relations
\begin{eqnarray}
(nm+2n+m+1) c_{n+1,m}&=&(m+1)(m+2)d_{n,m+2}
\label{eq:UI89DSAIT}\\
(n+1)(n+2) c_{n+2,m-2}&=&(nm-n+m)d_{n+1,m}
\label{eq:UI89ZE7T2}
\mbox{}
\end{eqnarray} 
to determine  $c_{n,m}$ and $d_{n,m}$. These relations may be used to 
calculate those values of $n,\,m$ which allow for non-vanishing coefficients 
and to calculate the proportionality constants between these coefficients; 
e.g. Eq.~\eqref{eq:UI89DSAIT} may be used to express $d_{n,m+2}$  in terms 
of $c_{n+1,m}$ provided $m \neq -1$ and $m \neq -2$. One obtains the result that 
the general solution $\beta(\rho,\rho^{\prime},\rho^{\prime \prime})$ of~\eqref{eq:DWLFF98DX} of polynomial 
form is given by~\eqref{eq:IB23AMF7F}, with   
\begin{eqnarray}
C(\rho,\rho^{\prime})&=&\sum_{n \in \mathcal{I}} 
C_{n}\rho^{n} (\rho^{\prime})^{-n}
\label{eq:UCCE3IZT}\\
D(\rho,\rho^{\prime})&=& A\rho - \sum_{n \in \mathcal{I}}
\frac{n-1}{n-2}
\,C_{n}\, \rho^{n-1} (\rho^{\prime})^{-n+2}
\label{eq:UDDE38ZT}
\mbox{,}
\end{eqnarray}   
where $A$ and $C_{n}$ are arbitrary constants and the 
index set $\mathcal{I}$ is given by 
\begin{equation}
  \label{eq:HEE12EIS}
\mathcal{I} = \left\{ n | n \in \mathcal{Z}, n\leq 0, n \geq 3 \right\} 
\mbox{.}
\end{equation}
While the derivation of~\eqref{eq:UCCE3IZT},~\eqref{eq:UDDE38ZT} is 
straightforward but lengthy, the fact 
that~\eqref{eq:IB23AMF7F}, \eqref{eq:UCCE3IZT},~\eqref{eq:UDDE38ZT}
fulfills~\eqref{eq:DWLFF98DX} may be verified easily. 

At this point we are looking for further constraints in order to 
reduce the number of unknown constants. The simplest (nontrivial) 
special case 
of~\eqref{eq:IB23AMF7F},~\eqref{eq:UCCE3IZT},~\eqref{eq:UDDE38ZT} is 
$C_{n}=0,\,\forall n \in \mathcal{I}$. The corresponding solution for 
$L_{0}=\beta/\rho$ is given by $L_{0}=A$. However, a solution given by a 
nonzero constant $A$ may be eliminated by adding a corresponding 
constant to the potential in $L(x,t)$. Thus, this solution need not 
be taken into account and we may set $A=0$.     

The 'next simplest' solution, given by $A=0$, $C_{n}=0$ for all 
$n \in \mathcal{I}$ except $n=0$, takes the form
\begin{equation}
  \label{eq:TNS22ST9SC}
L_{0}= L_{0}^{(0)}=C_0
\left( 
\frac{\rho^{\prime \prime}}{\rho}
-\frac{1}{2}\frac{(\rho^{\prime})^{2}}{\rho^{2}}
\right)=
2\,C_0 \frac{1}{\sqrt{\rho}}\frac{\partial^{2}\sqrt{\rho}}{\partial x^{2}}
\mbox{.}
\end{equation}
Let us also write down here, for later use, the solution given by 
$A=0$, $C_{n}=0$ for all $n \in \mathcal{I}$ except $n=-1$. It takes  
the form
\begin{equation}
  \label{eq:NEUEL89Q2}
L_{0}= L_{0}^{(-1)}=C_{-1}
\left( 
\frac{\rho^{\prime}}{\rho^{2}} \rho^{\prime \prime}
-\frac{2}{3}\frac{(\rho^{\prime})^{3}}{\rho^{3}}
\right)
\mbox{,}
\end{equation}

Comparison of the r.h.s. of Eq.~\eqref{eq:TNS22ST9SC} with 
Eq.~\eqref{eq:DZMFE8NEU} shows that the solution~\eqref{eq:TNS22ST9SC} 
leads to Schr{\"o}dinger's equation~\eqref{eq:SCH43STRD}. At this point
the question arises why this particular solution has been realized by 
nature - and not any other from the huge set of possible solutions.   
Eq.~\eqref{eq:TNS22ST9SC} consists of two parts. Let us consider the 
two corresponding terms in $\rho L_0$, which represent two contributions 
to the Lagrangian in Eq.~\eqref{eq:ANKDRDDC}. The second of these terms 
agrees with the integrand of the Fisher functional~\eqref{eq:fisherfun}. 
The first is proportional to $\rho^{\prime \prime}$. This first term 
may be omitted in the Lagrangian (under the integral sign) 
because it represents a boundary 
(or surface) term and gives no contribution to the Euler-Lagrange 
equations [it must \emph{not} be omitted in the final differential 
equation~\eqref{eq:DVGDG8U}, where exactly the same term reappears as 
a consequence of the differentiation of $\rho^{\prime}$]. Thus, 
integrating the 
contribution $\rho L_0$ of the solution~\eqref{eq:TNS22ST9SC} to the 
Lagrangian yields exactly the Fisher functional. No other solution 
with this property exists. Therefore, the reason why nature has chosen 
this particular solution is basically the same as in classical statistics, 
namely the principle of maximal disorder - but realized in a different 
(local) context and expressed in terms of a principle of minimal Fisher 
information. 

We see that the conflict mentioned at the beginning of this section 
does not exist for the quantum mechanical solution~\eqref{eq:TNS22ST9SC}.
The reason is again that the term in $L_0$ containing the second derivative 
$\rho^{\prime \prime}$ is of the form of a total derivative and can, 
consequently, be 
neglected as far as its occurrence in the term $\rho L_0$ of the Lagrangian
is concerned. Generalizing this fact, we may formulate the following 
criterion for the absence of any conflict: The terms in $L_0$ containing 
$\rho^{\prime \prime}$ must not yield contributions to the variation, i. e. 
they must 
in the present context take the form of total derivatives (for more 
general variational problems such terms are called 
"`null Lagrangians"'~\cite{giaquinta.hildebrandt:variations1}).      

So far, in order to reduce the number of our integration constants, we used 
the criterion that the corresponding term in the Lagrangian should agree 
with the form of Fishers functional. This `direct' implementation of the 
principle of maximal disorder led to quantum mechanics. The absence of 
the above mentioned conflict means that the theory may be formulated 
using a Lagrangian containing no derivatives higher than first order. 
As is well known, this is a criterion universally realized in nature; a 
list of fundamental physical laws obeying this criterion may e.g. be 
found in a paper by Frieden and Soffer~\cite{frieden_soffer:lagrangians}. 
Thus, it is convincing although of a 'formal' character. Let us apply 
this 'formal' criterion as an alternative physical argument to reduce 
the number of unknown coefficients the above solution. This criterion 
implies, that the derivatives of $L_0$ with respect 
to $\rho^{\prime \prime}$ do not play any role, i.e. the 
solutions of~\eqref{eq:DWLFF98DX} must also obey
\begin{equation}
  \label{eq:DWJZ7WDX}
\frac{\mathrm{d}}{\mathrm{d}x}
\frac{\partial \beta}{\partial \rho^{\prime}}
-\frac{\partial \beta}{\partial \rho}
+\frac{\beta}{\rho}=0
\mbox{.}
\end{equation}
This implies that only those solutions of~\eqref{eq:DWLFF98DX} are 
acceptable, which obey 
\begin{equation}
  \label{eq:DWLKQW34E}
-\frac{\mathrm{d}^{2}}{\mathrm{d}x^{2}}
\frac{\partial \beta}{\partial \rho^{\prime \prime} }=0
\mbox{.}
\end{equation} 
Using~\eqref{eq:IB23AMF7F} and~\eqref{eq:UCCE3IZT} it is 
easy to see that the solution~\eqref{eq:TNS22ST9SC} belonging 
to $n=0$ is the only solution compatible with the 
requirement~\eqref{eq:DWLKQW34E} [as one would suspect it is also
possible to derive~\eqref{eq:TNS22ST9SC} directly 
from~\eqref{eq:DWJZ7WDX}]. Thus the 'formal' principle, that 
the Lagrangian contains no terms of order higher than one, leads
to the same result as the 'direct' application of the principle
of maximal disorder. The deep connection between statistical criteria
and the form of the kinetic energy terms in the fundamental laws of 
physics has been mentioned before in the 
literature~\cite{frieden:sciencefisher}. The present derivation 
sheds new light, from a different perspective, on this connection.  

Summarizing, the shape of our unknown function $L_0$ has been found. 
The result for $L_0$ leads to Schr{\"o}dinger's equation, as pointed out 
already in section~\ref{sec:statisticaltheories}. This means that quantum 
mechanics may be selected from an infinite set of possible theories 
by means of a logical principle of simplicity, the statistical 
principle of maximal disorder. Considered from this point of view quantum 
mechanics is 'more reasonable' than its classical limit (which is 
a statistical theory like quantum mechanics). It also means (see 
section~\ref{sec:energyconservation}) that the choice $h=0$ is 
justified as far as the calculation of expectation values of 
$p^{n}, n\le{2}$ is concerned.   

In closing this section we note that the particular form of the 
function $L(x,t)$ has never been used. Thus, while the calculation 
of $L_0$ reported in this section completes our derivation of 
quantum mechanics, the result obtained is by no means specific 
for quantum mechanics. Consider the steps leading from 
the differential equation~\eqref{eq:DVGDG8U} and 
the variational principle~\eqref{eq:AJ8UZ4DC} to the general 
solution~\eqref{eq:IB23AMF7F},~\eqref{eq:UCCE3IZT},~\eqref{eq:UDDE38ZT}.
If we now supplement our previous assumptions with the composition 
law~\eqref{eq:ASCLIAA}, we are able to single out the Fisher $I$ 
among all solutions [compare e.g.~\eqref{eq:NEUEL89Q2} 
and~\eqref{eq:TNS22ST9SC}]. Thus, the above calculations may also 
be considered as a new derivation of the Fisher functional, based 
on assumptions different from those used previously in the literature. 

\section{Discussion}
\label{sec:discussion}
Both the formal transition from classical physics to quantum mechanics
(quantization procedure) and the interpretation of the resulting 
mathematical formalism is presently dominated by the particle picture. 

To begin with the interpretation, Schr{\"o}dinger's 
equation is used to describe, e.g., the behavior of \emph{individual} 
electrons. At the same time the \emph{statistical} nature of quantum 
mechanics is obvious and cannot be denied. To avoid this fundamental 
conflict, various complicated intellectual constructions, which I do 
not want to discuss here, have been - and are being- designed. But 
the experimental data from the micro-world (as interpreted in the 
particle picture) remain mysterious, no matter which one of these 
constructions is used.

Let us now consider the quantization process. The canonical quantization 
procedure consists of a set of formal rules, which include, in particular, 
the replacement of classical momentum and energy observables $p,E$ by new 
quantities, according to   
\begin{equation}
  \label{eq:DTR43QUPR}
p\to\frac{\hbar}{\imath}\frac{\mathrm{d}}{\mathrm{d}x},\;\;\;
E\to-\frac{\hbar}{\imath}\frac{\mathrm{d}}{\mathrm{d}t},\;\;\;
\mbox{,}
\end{equation}
which then act on states of a Hilbert state, etc. By means of this 
well-known set of rules one obtains immediately Schr{\"o}dinger's 
equation~\eqref{eq:SCH43STRD} from the classical Hamiltonian of a single 
particle. While we are accustomed to 'well-established' rules 
like~(\ref{eq:DTR43QUPR}), it is completely unclear why they work. It 
does not help if more sophisticated  versions of the canonical quantization 
procedure are used. If for example, the structural similarity between 
quantum mechanical commutators and classical Poisson 
brackets~\cite{sudarshan:classical} is used as a starting 
point, this does not at all change the mysterious nature of the 
jump into Hilbert space given by~\eqref{eq:DTR43QUPR}; this structural 
similarity is just a consequence of the fact that both theories share 
the same space-time (symmetries). 

Thus, there seems to be no possibility to understand either the 
quantization procedure or the interpretation of the formalism, if a 
single particle picture is used as a starting point. According to 
a (prevailing) positivistic attitude this is no problem, since the 
above rules 'work' [they illustrate perfectly von Neumann's saying 
"`In mathematics you don't understand things. You just get used to 
them"']. On the other hand, an enormous amount of current activity 
\emph{is} apparently aimed at an understanding of quantum mechanics. 

We believe that the particle picture is inadequate, both as a starting 
point for the quantization process and with regard to the interpretation
of the formalism. In fact, both of these aspects seem to be intimately 
related to each other;  a comprehensible quantization procedure will 
lead to an adequate interpretation and a reasonable quantization 
procedure will only be found if the theory is interpreted in an 
adequate way. The position adopted here is that the most adequate way is 
the \emph{simplest} possible way. We believe that quantum mechanics 
is a statistical theory whose dynamical predictions make only sense for 
statistical ensembles and cannot be used to describe the behavior of 
individual events. This ensemble interpretation of quantum mechanics 
(for more details the reader is referred to review articles by 
Ballentine~\cite{ballentine:statistical} and by Home and 
Whitaker~\cite{home:ensemble}) is generally accepted as the simplest 
possible interpretation, free from any contradictions and free from 
any additional assumptions expanding the range of validity of the 
original formalism. Fundamental conceptual problems like the 
``measurement problem'' or the impossibility to characterize the 
wave function of a single particle by means of experimental 
data~\cite{alter:quantum} do not exist in the statistical 
interpretation. Nevertheless it is a minority view; the reason 
may be that it forces us to accept that essential parts of reality 
are out of our control. This inconvenient conclusion can be 
avoided by \emph{postulating} that all fundamental laws of nature 
must be deterministic (with regard to the description of individual 
events). From the point of view of this deterministic dogma, any 
interpretation denying the completeness of QM  must be a 
``hidden variable theory''.

If we accept the ensemble interpretation of quantum mechanics, then the 
proper starting point for quantization must be a statistical theory. The 
assumption of a Hilbert space for the considered system should be 
avoided. This would mean postulating many essential quantum mechanical 
properties without any possibility to analyze their origin. Our aim is 
the derivation of Schr{\"o}dinger's equation, from which then, afterwards, 
the Hilbert space structure can be obtained by means of mathematical 
analysis, abstraction and generalization~\cite{messiah:quantum*1}. 
Preferably, Schr{\"o}dinger's equation should be derived from assumptions 
which can be understood in the framework of general classical 
(statistical as well as deterministic) and logical concepts. This 
route to quantum mechanics is of course not new. Any listing of 
works~\cite{schrodinger:quantisierung_I,motz:quantization,schiller:quasiclassical,rosen:classical_quantum,frieden:fisher_basis,reginatto:derivation,lee.zhu:principle,hall.reginatto:schroedinger,hall.reginatto:quantum_heisenberg,frieden:sciencefisher,nikolic:classical,syska:fisher,klein:schroedingers}
following related ideas must necessarily be incomplete. In the present 
paper an attempt has been undertaken to find a set of assumptions which 
is on the one hand \emph{complete} and on the other hand as 
\emph{simple and fundamental} as possible. Throughout this work 
all calculations have been performed for simplicity for a single spatial 
dimension. In the meantime, after submission of this paper, the present 
approach has been generalized to three dimensions, gauge 
fields, and spin~\cite{klein:gauge_spin}. Given that it can 
be further generalized to a $3N$-dimensional configuration 
space, this would mean that essentially all of non-relativistic 
quantum mechanics can be derived in the framework of the present 
approach (preliminary calculations of the present author indicate 
that this can indeed be done). This aim has not yet been completely 
achieved but I will sometimes tacitly assume in the following discussion 
that it can be achieved. 

Our first, and - in a sense - central assumption was the set of 
relations~(\ref{eq:FIRSTAET}),~(\ref{eq:SECONAET}), which may be 
characterized as a statistical version of the two fundamental equations 
of classical mechanics displayed in~(\ref{eq:FSTASECCANEQ}), namely the 
definition of particle momentum and Newton's equation. In writing 
down these relations the existence of two random variables $x$ and $p$, 
with possible values from $\mathcal{R}$, has been postulated. This 
means that appropriate experimental devices for measuring position and 
momentum may be set up. The probabilities $\rho(x,t)$ and $w(p,t)$ 
are observable quantities, to be determined by means of a large number of 
individual measurements of $x$ and $p$. Given such data for $\rho(x,t)$ 
and $w(p,t)$ the validity of the statistical 
conditions~(\ref{eq:FIRSTAET}),~(\ref{eq:SECONAET}) may be tested.
Thus, these relations have a clear operational meaning. On the 
other hand, they do not provide a statistical law of nature, i.e. 
dynamical equations for the probabilities $\rho(x,t)$ and $w(p,t)$. 
Further constraints are required to define such laws. As shown in 
section~\ref{sec:statisticaltheories} the statistical framework 
provided by~(\ref{eq:FIRSTAET}),(\ref{eq:SECONAET}) is very general; 
it contains quantum mechanics and its classical limit as well 
as an infinite number of other theories. Thus, it provides a 
"`bird's eye view"' on quantum theory.

Our second postulate was the validity of a continuity equation
of the form~(\ref{eq:CONTONEDIM}). For the type of theory considered 
here the validity of a local conservation law of probability is a 
very weak assumption - more or less a logical necessity. The 
special form of the probability current postulated 
in~(\ref{eq:CONTONEDIM}) is suggested by Hamilton-Jacobi theory 
(this is of course only an issue for spatial dimensions higher than one). 
It means that an ensemble of particles is considered for which a 
\emph{wave front} may be defined. A detailed study of such sets of 
particle trajectories, which are referred to as "`coherent systems"', 
may be found in a review article by Synge~\cite{synge:classical}.
In fluid dynamics~\cite{faber:fluid_dynamics} corresponding 
fields are called "`potential flow fields"' 

The first two postulates led to an infinite number of statistical 
theories (coupled differential equations for  $\rho$ and $S$) 
characterized by an unknown term $L_{0}$. Only the classical (limit) 
theory, defined by $L_{0}=0$, allows for an identification of objects 
independent from the statistics; in this case the differential equation 
for $S$ does not depend on $\rho$. In all other (non-classical) theories 
there is a $\rho$-dependent coupling term preventing such an 
identification.

Our third postulate was the assumption that the remaining unknown 
function of $\rho$ and  $\rho^{\prime}$, in the coupled differential 
equations for  $\rho$ and $S$, takes a form which is in 
agreement with the principle of maximal disorder (or minimal knowledge). 
This logical principle of simplicity is well known, in the form of a 
postulate of maximal entropy, from statistical thermodynamics. In the 
present case it has to be implemented in a different and more complicated 
way, as a postulate of minimal Fisher information. This is due to the fact 
that the present equation for the determination of $\rho$ is a differential
equation, i.e. it may not only depend on $\rho$ but also on derivatives 
of $\rho$. The entropy is a functional which depends only on $\rho$ and 
is unable to adjust properly with regard to this new 'degree of freedom'. 
Our analysis started in section~\ref{sec:disorder} with a discussion
of the conventional principle of maximal entropy and led to the variational 
principle~(\ref{eq:AJ8UZ4DC}) in section~\ref{sec:subsidiary-condition}, 
which formally describes the principle of maximal disorder in the present 
context. Finally, in section~\ref{sec:derivationofL}, 
Eq.~(\ref{eq:AJ8UZ4DC}) has been used to determine the unknown term, 
which leads to Schr{\"o}dinger's equation.

Schr{\"o}dinger's equation says nothing about the calculation of expectation 
values of $p-$dependent quantities. To reproduce this part of the 
quantum mechanical formalism, we had to implement a further requirement, 
namely conservation of energy in the mean. From this fourth assumption 
the standard quantum mechanical result could be recovered for terms of 
the form $p^{n}$, where $n=0,1,2$. These terms are not the only ones 
occurring in realistic situations. If we want to study the behavior 
of charged particles in a magnetic field we should be able to calculate 
expectation values of terms of the form $ps(x)$, where $s(x)$ is an 
arbitrary function of $x$. Such terms (and more generally the inclusion 
of gauge fields) will be dealt with in future work.     
  
Summarizing, the most important relations of quantum mechanics have 
been derived from assumptions which may be characterized either as 
purely statistical or as statistical versions (or continuum versions) 
of relations of particle physics. The continuity equation and the 
principle of maximal disorder belong to the former class. The 
statistical conditions, conservation of energy in the mean, and the 
special form of the probability current belong to the latter class.
These statistical assumptions imply quantum mechanics and are much
simpler to understand than the jump into Hilbert space given by 
Eq.~(\ref{eq:DTR43QUPR}). Of course, all of these assumptions are relations 
or structural properties belonging to the quantum-mechanical formalism;
it would not be possible to derive quantum mechanics from assumptions
which are \emph{not} quantum-mechanical in nature. However, it is not 
trivial that these, relatively simple and comprehensible assumptions are 
sufficient to derive the basic relations of the whole formalism.

The above derivation of the most basic equations of quantum 
mechanics from statistical assumptions presents a strong argument in 
favor of a statistical (ensemble) interpretation. This becomes even 
more evident, if the relation between quantum mechanics and its 
classical limit is considered in detail. As discussed in 
section~\ref{sec:randomvariables} the transition from the classical 
to the quantum mechanical theory is characterized by the elimination 
of a deterministic element, namely the (deterministic) functional 
relation between position and momentum variables. Thus, quantum 
mechanics contains less deterministic elements (it is 
'more statistical' in nature) than its classical limit, a result 
in accordance with the general classification scheme set up in 
section~\ref{sec:onprobability}. This loss of determinism is implicitly 
contained in the above assumptions and presents the essential 
"`non-classical"' element of the present derivation. It is 
interesting to compare the present derivation with other 
derivations of Schr{\"o}dinger's equation making use of 
different 
"`non-classical 
elements"'~\cite{hall.reginatto:quantum_heisenberg,lee.zhu:principle}.  
The loss of determinism mentioned above is also responsible for the 
crucial role of the concept of Fisher information in the present 
work. This concept was realized here in a way different from the one 
followed previously by Frieden and others~\cite{frieden:sciencefisher,frieden_soffer:lagrangians,hall.reginatto:quantum_heisenberg}. On the other hand, 
several aspects of the present work may also be seen as a complement 
to this previous approach. In particular, the 'classical limit theory' 
which is used as a starting point by these authors may be 
derived from the first two assumptions of the present work.

Many works on the foundations of quantum mechanics are motivated by 
the wish to identify deterministic,  or at least 'classical-probabilistic' 
elements in its structure. According to the original point of view 
of Schr{\"o}dinger, he derived a classical wave 
equation and $|\psi|^{2}$ was (in contrast to the present 
interpretation as a probability density) an \emph{observable} 
field measuring something like the density distribution of an 
'extended particle'. More recently, this interpretation has been 
reconsidered in an interesting paper by Barut~\cite{barut:combining}. 
But the results of modern high-precision 
measurements~\cite{tonomura:demonstration} strongly support the 
probabilistic interpretation and exclude, in our opinion, this 
original classical view of Schr{\"o}dinger. 

Since it turned out that probability plays an indispensable role in 
quantum theory, various attempts have been undertaken to interpret it 
(at least) as a theory with classical randomness. A well-known example 
is Nelson's stochastic mechanics~\cite{nelson:quantum}, where a 
stochastic background field of unspecified origin, combined with 
deterministic mechanics, leads to Schr{\"o}dinger's equation. There 
is a certain overlap of ideas between such theories and the present 
one; in both cases a probabilistic equation is derived from a 
set of well-defined (partly) probabilistic assumptions. However, the 
present theory contains neither particle trajectories nor stochastic 
forces. Probability is introduced in a more abstract way by means 
of a postulated conservation law and the occurrence of expectation values.
in the basic equations defining the theory.

Finally, a theory by Khrennikov~\cite{khrennikov:prequantum_classical},~\cite{khrennikov:prequantum_complex} should be mentioned which continues and extends both 
the stochastic approach and the original Schr{\"o}dinger point of view, with 
the aim of reconciling the latter with the probability concept. To achieve 
this goal, classical fluctuating fields in an infinite-dimensional phase 
space are introduced by means of appropriate mathematical axioms. Despite 
the completely different language, this work is basically written in the 
same spirit as stochastic theories. As for a comparison with the present 
theory similar remarks as above apply. 

Despite the similarities mentioned above, the methodic basis 
of the present work differs considerably from the stochastic 
approach. Indeterminism, as regards individual events in the 
micro-world, is considered as an irreducible feature of nature. 
This position is not incompatible with the fact that macroscopic 
bodies follow deterministic laws; these bodies are aggregates 
of a large number  $N$ of individual particles and it is 
reasonable (looking at examples from many-body physics) to 
expect that all quantum uncertainties will be somehow 
'washed out' for large $N$; a similar problem has been analyzed 
in a remarkable paper by Cini and Serva~\cite{cini_serva:where}. 
Classical physics, as it is generally understood (neglecting 
back-reaction from self-fields, see~\cite{klein:schroedingers}
for more details), contains \emph{less} uncertainty as compared 
to quantum mechanics. It is therefore impossible to derive 
quantum mechanics from classical physics, just as it would be 
impossible to derive classical statistical physics from 
thermodynamics.

\section{Concluding remarks}
\label{sec:concludingremarks}

The present derivation was based on the assumption that dynamical 
predictions are only possible for statistical averages and not for 
single events - leaving completely open the question \emph{why} 
predictions on single events are impossible. This deep question 
remains unanswered; some speculative remarks on a possible source 
of the indeterminacy have been given elsewhere~\cite{klein:schroedingers}. 

As regards the generalization of the present quantization method to fields, two 
different routes seem feasible. The first makes use of the well-known fact 
that the formalism of second quantization may be deduced, in the 
limit $N \to \infty$, from the N-particle Schr{\"o}dinger 
equation~\cite{fetter:many_particle}. Thus, proceeding along this 
route, the decisive step is the derivation of the many-particle 
Schr{\"o}dinger equation. This can be done in the present approach; 
all assumptions can be generalized in a most natural way (just increasing 
the number of variables) to cover the many-body situation. The second route 
starts from classical fields and applies the usual quantization rules; but now 
for an infinite number of degrees of freedom.  In order to generalize the 
present approach in an analogous way, the quantity $\rho$ must be interpreted 
in a different way, as a density of a stream of particles in the framework 
of an approximate continuum theory. This $\rho$ would be an 
observable quantity in the sense of Schr{\"o}dinger and de Broglie. 
In this respect the resulting differential equation, which is 
again~\eqref{eq:SCH43STRD}, should be considered as a classical field 
equation - despite the presence of a parameter $\hbar$ (solutions of this 
pre-quantum Schr{\"o}dinger equation have been studied in the 
literature~\cite{fick:einfuehrung}). However, it would be  
only \emph{approximately} true - what is actually observed 
are particles and not fields. This would provide a motivation for a 
more accurate description (second quantization). This second route 
to field quantization presents an open problem for future research. 
It might be interesting in view of some conceptual problems of 
quantum field theory. 

'Interaction between individual objects' and the corresponding notion of 
\emph{force} are macroscopic concepts. In the microscopic domain, where
according to the present point of view only statistical laws are valid, the 
concept of force looses its meaning. In fact, in the quantum-mechanical 
formalism 'interaction' is not described in terms of forces but in terms 
of \emph{potentials} (as is well known, this leads to a number of subtle 
questions concerning the role of the vector potential in quantum mechanics). 
The relation between these two concepts is still not completely understood;
the present statistical approach offers a new point of view to study this 
problem~\cite{klein:gauge_spin}.

In a previous work~\cite{klein:schroedingers} of the present author, 
Schr{\"o}dinger's equation has been derived from a different set of 
assumptions including the postulate that the dynamic equation of 
state may be formulated by means of a complex-valued state 
variable $\psi$. The physical meaning of this assumption is 
unclear even if it sounds plausible from a mathematical point 
of view. The present paper may be seen as a continuation and 
completion of this previous work, insofar as this purely mathematical 
assumption has been replaced now by other requirements which may be 
interpreted more easily in physical terms. Finally, we mention
that the present theory has already been generalized to three spatial 
dimensions, gauge fields, and spin~\cite{klein:gauge_spin}. Important 
open questions for future research, extending the range of validity 
of the present approach, include a higher-dimensional configuration 
space (several particles), a relativistic formulation, and an 
extension of the present method to fields.

\bibliography{uftbig}

\end{document}